\documentclass{aastex631}
\usepackage{amssymb,amsmath}

\shorttitle{ASTERIA - Asteroid Thermal Inertia Analyzer}
\shortauthors{Novakovi\'c et al.}
\begin{document}

\title{ASTERIA - Asteroid Thermal Inertia Analyzer}

\author[0000-0001-6349-6881]{Bojan Novakovi\' c}
\affiliation{Instituto de Astrofísica de Andalucía (CSIC) \\
Glorieta de la Astronomía s/n \\
E-18080 Granada, Spain}
\affiliation{Department of Astronomy, Faculty of Mathematics, University of Belgrade \\
Studentski trg 16 \\
11000 Belgrade, Serbia}

\author[0000-0002-7058-0413]{Marco Fenucci}
\affiliation{ESA ESRIN / PDO / NEO Coordination Centre \\
Largo Galileo Galilei, 1 \\
00044 Frascati (RM), Italy}
\affiliation{Elecnor Deimos \\
Via Giuseppe Verdi, 6 \\
28060 San Pietro Mosezzo (NO), Italy}
\affiliation{Department of Astronomy, Faculty of Mathematics, University of Belgrade \\
Studentski trg 16 \\
11000 Belgrade, Serbia}

\author[0000-0003-4706-4602]{Du\v{s}an Mar\v{c}eta}
\affiliation{Department of Astronomy, Faculty of Mathematics, University of Belgrade \\
Studentski trg 16 \\
11000 Belgrade, Serbia}

\author[0000-0001-7008-4367]{Debora Pavela}
\affiliation{Department of Astronomy, Faculty of Mathematics, University of Belgrade \\
Studentski trg 16 \\
11000 Belgrade, Serbia}

%% Note that the \and command from previous versions of AASTeX is now
%% depreciated in this version as it is no longer necessary. AASTeX 
%% automatically takes care of all commas and "and"s between authors names.

%% AASTeX 6.31 has the new \collaboration and \nocollaboration commands to
%% provide the collaboration status of a group of authors. These commands 
%% can be used either before or after the list of corresponding authors. The
%% argument for \collaboration is the collaboration identifier. Authors are
%% encouraged to surround collaboration identifiers with ()s. The 
%% \nocollaboration command takes no argument and exists to indicate that
%% the nearby authors are not part of surrounding collaborations.

%% Mark off the abstract in the ``abstract'' environment. 
\begin{abstract}

Thermal inertia estimates are available for a limited number of a few hundred objects, and the results are practically solely based on thermophysical modeling (TPM). 
We present a novel thermal inertia estimation method, Asteroid Thermal Inertia Analyzer (ASTERIA). The core of the ASTERIA model is the Monte Carlo approach, based on the Yarkovsky drift detection. We validate our model on asteroid Bennu plus ten well-characterized near-Earth asteroids (NEAs) for which a good estimation of the thermal inertia from the TPM exists. The tests show that the ASTERIA provides reliable results consistent with the literature values.
The new method is independent from the TPM, allowing an independent verification of the results. As the Yarkovsky effect is more pronounced in small asteroids, the noteworthy advantage of the ASTERIA compared to the TPM is the ability to work with smaller asteroids for which TPM typically lacks the input data. We used the ASTERIA to estimate the thermal inertia of 38 NEAs, with 31 of them being sub-km asteroids. Twenty-nine objects in our sample are characterized as Potentially Hazardous Asteroids.
On the limitation side, the ASTERIA is somewhat less accurate than the TPM. The applicability of our model is limited to NEAs, as the Yarkovsky effect is yet to be detected in main-belt asteroids. However, we can expect a significant increase in high-quality measurements of the input parameters relevant to the ASTERIA with upcoming surveys. This will surely increase the reliability of the results generated by the ASTERIA and widen the model's applicability.

\end{abstract}

%% Keywords should appear after the \end{abstract} command. 
%% The AAS Journals now uses Unified Astronomy Thesaurus concepts:
%% https://astrothesaurus.org
%% You will be asked to selected these concepts during the submission process
%% but this old "keyword" functionality is maintained in case authors want
%% to include these concepts in their preprints.

%% From the front matter, we move on to the body of the paper.
%% Sections are demarcated by \section and \subsection, respectively.
%% Observe the use of the LaTeX \label
%% command after the \subsection to give a symbolic KEY to the
%% subsection for cross-referencing in a \ref command.
%% You can use LaTeX's \ref and \label commands to keep track of
%% cross-references to sections, equations, tables, and figures.
%% That way, if you change the order of any elements, LaTeX will
%% automatically renumber them.
%%
%% We recommend that authors also use the natbib \citep
%% and \citet commands to identify citations.  The citations are
%% tied to the reference list via symbolic KEYs. The KEY corresponds
%% to the KEY in the \bibitem in the reference list below. 

\section{Introduction}
\label{sec:intro}
Despite their great importance, knowledge of the physical properties of asteroids lags far behind the current rate of their discoveries. To better understand asteroids' properties, one of the most critical parameters to estimate is their surface thermal inertia. For instance, the value of thermal inertia can give essential constraints on the type of material present on the surface. However, as the asteroid's thermal inertia varies in a considerable interval of values spanning more than an order of magnitude, it cannot be reliably assumed a priori. Yet, the determination of the surface thermal inertia is a challenging step. 

Data provided by space missions to asteroids and comets are extremely valuable information. Among them, NASA's Dawn space probe visited asteroids Vesta and Ceres \citep{2015aste.book..419R}, while ESA's Rosetta mission performed a detailed study of comet 67P/Churyumov–Gerasimenko \citep{2007SSRv..128....1G}. The most recent examples are two robotic sample-return projects: JAXA's Hayabusa2 and NASA's OSIRIS-REx, exploring Ryugu and Bennu, respectively \citep[see, e.g.][]{watanabe-etal_2019, lauretta-etal_2019}. The level of detail, reliability, and precision of the thermal inertia values provided by these missions is outstanding \citep[see, e.g.][]{2014GeoRL..41.1438C,2020JGRE..12505733R,grott-etal_2020,2020SciA....6.3699R}. However, such data is available only for a small sample of objects, and it is difficult to extrapolate these results to other objects or to derive global properties valid to a population level from a small set of well-known asteroids. Nevertheless, the spacecraft visits provided data that are key for testing and validating the other models used for thermal inertia estimations.

Typically, the thermal inertia is estimated by thermophysical models, making use of thermal infrared observations \citep[see, e.g.][and references therein]{delbo-etal_2015}. Such observations, however, are available for a limited number of asteroids. Until several years ago, the thermal inertia was estimated for a very small number of asteroids \citep{delbo-etal_2015}. The situation started to change recently, primarily thanks to the Wide-field Infrared Survey Explorer \citep[WISE,][]{2010AJ....140.1868W} and its post-cryogenic survey extension known as NEOWISE \citep{2014ApJ...792...30M}.
\citet{2020AJ....159..264J} investigate the size, thermal inertia, surface roughness, and geometric albedo of 10 Vesta family asteroids using the Advanced Thermophysical Model based on the thermal-infrared data acquired by WISE. Similarly, \citet{2021AJ....162...40J} derived thermal inertia for 20 Themis asteroid family members from thermophysical modeling based on the WISE/NEOWISE observations. 
\citet{2021PSJ.....2..161M} used infrared data for a set of 239 objects observed by the WISE to estimate their size, albedo, thermal inertia, and surface roughness.
\citet{2021A&A...654A..87M} used the light curves datasets in combination with the thermal infrared data, mainly from the Infrared Astronomical Satellite \citep[IRAS,][]{1984ApJ...278L...1N}, AKARI (aka as ASTRO-F or IRIS - InfraRed Imaging Surveyor) \citep{2011PASJ...63.1117U}, and WISE in an optimization process using the Convex Inversion Thermophysical Model. The authors analyzed the properties of 16 slowly rotating asteroids, including estimating their thermal inertia.
\citet{2022PSJ.....3...56H} used thermal flux measurements obtained by the WISE/NEOWISE during its fully cryogenic phase to obtain new thermophysical model fits of 1847 asteroids, deriving their thermal inertia, diameter, and Bond and geometric albedos.
\citet{2023ApJ...944..202J} studied the surface properties of asteroid (704) Interamnia, combining the mid-infrared wave bands observations from the Subaru telescope along with those obtained by 
IRAS, AKARI, and WISE/NEOWISE. 
\citet{2020A&A...638A..84A} used Herschel Photodetector Array Camera and Spectrometer data taken during the Asteroid Preparatory Programme for Herschel, ASTRO-F and the Atacama Large Millimeter/submillimeter Array \citep{2005ESASP.577..471M} for a thermo-physical characterization of 18 large main belt asteroids.  

We briefly reviewed only the most recent works as they provide the most currently available thermal inertia estimates. Nevertheless, we still have thermal inertia estimations for only a few hundred asteroids, primarily large main-belt asteroids, and some near-Earth asteroids (NEAs). Therefore, it is still an essential task, and the surface thermal inertia (TI) estimation for any new object is very important. Alternative methods for the TI estimation, using other kinds of observations, would be of great scientific value.  

Recently, \citet{2021A&A...647A..61F} introduced a statistical method for the thermal estimation of NEAs based mainly on ground-based observations. The method was successfully applied to determine the thermal inertia of two rapidly rotating NEAs, asteroids 2011~PT and 2016~GE1 \citep{2021A&A...647A..61F,Fenucci_et_al_AA2023}.
The method requires, in fact, a determination of the semi-major axis drift produced by the Yarkovsky effect \citep[see, e.g.][for a review]{bottke-etal_2006, vokrouhlicky-etal_2015}, a thermal effect caused by sunlight that strongly depends on the surface properties. Measurements of the Yarkovsky effect can be performed through orbit determination employing astrometric and/or radar observations \citep[][]{chesley-etal_2003, milani-gronchi_2009}, which are typically ground-based. Then, by matching the measured semi-major axis drift with that predicted by a physical model, the method produces a distribution of the surface thermal inertia $\Gamma$, possibly giving further constraints on other physical parameters. 
The Yarkovsky effect has been detected for hundreds of NEAs \citep[][]{farnocchia-etal_2013, 2018A&A...617A..61D, greenberg-etal_2020}, and more and more detections are foreseen in the future because of two reasons: 1) the length of the observational arc can be only extended when NEAs are re-observed, and 2) new ground-based surveys like the Vera Rubin Observatory \citep{2018Icar..303..181J} and the ESA Flyeye telescope will provide more chances of re-observe known NEAs or recover the lost ones. The determination of the Yarkovsky effect is also starting to become a routine operation for the automatized orbit determination systems provided by the ESA NEO Coordination Centre\footnote{\url{https://neo.ssa.esa.int/}} (NEOCC), the NEODyS\footnote{\url{https://newton.spacedys.com/}} service, and the NASA JPL Small-Body Database\footnote{\url{https://ssd.jpl.nasa.gov/}}. This situation allows us to apply the method by \citet{2021A&A...647A..61F} to many NEAs. An essential advantage of this model is that it can estimate the thermal inertia of comparatively smaller asteroids than the other methods. However, despite being successfully applied for rapidly rotating objects, the method proposed by \citet{2021A&A...647A..61F} still requires several improvements as well as additional validations before it can be safely applied to a large number of objects. 

Motivated by these facts, one of the aims of the "Demystifying Near-Earth Asteroids" (D-NEAs) project funded by The Planetary Society\footnote{\url{https://www.planetary.org/}} is to develop further methods for the characterization of NEAs based on ground-based observations. In this paper, we propose extensions of the concept introduced in \citet{2021A&A...647A..61F, Fenucci_et_al_AA2023} in order to make the thermal inertia estimation more reliable and accurate. In particular, we implemented different Yarkovsky effect models suitable for different dynamical and physical characteristics scenarios. Moreover, we refined the population-based modeling of poorly constrained or unknown physical parameters needed for the Yarkovsky modeling. We implemented all the methods and models described in the present paper in the Asteroid Thermal Inertia Analyzer (ASTERIA). The ASTERIA software v1.0.0 \citep{asteria_zenodo} is publicly available to the scientific community under the CC BY-NC-SA 4.0 license, and it can also be retrieved from a dedicated online repository\footnote{\url{https://github.com/Fenu24/D-NEAs}}.  

The paper is structured as follows. In Sec.~\ref{s:MC}, we describe the general idea and Monte Carlo method for thermal inertia estimation. In Sec.~\ref{s:yarkoModels}, we give the details of the Yarkovsky effect models included in the ASTERIA software. Sec.~\ref{s:physParam} is dedicated to the explanations on modeling of the physical parameters entering the Yarkovsky effect model. In Sec.~\ref{s:software}, we presented a basic introduction to the software, while Sec.~\ref{s:validation} is devoted to the model and code validation on Bennu and 10 other well-characterized NEAs. In Sec.~\ref{s:new_TI}, we have presented new thermal inertia estimates for 38 NEAs. Finally, we provide a summary and main conclusions in Sec.~\ref{s:summary}.

\section{Monte Carlo method of thermal inertia estimation}
\label{s:MC}
The method of thermal inertia estimation introduced by \citet{2021A&A...647A..61F} relies on the assumption that a measure $(\textrm{d}a/\textrm{d}t)_\textrm{m}$ of the Yarkovsky effect obtained from astrometric data \citep[see, e.g.][]{farnocchia-etal_2013, 2018A&A...617A..61D, greenberg-etal_2020} is available for the object we are interested in. Models for Yarkovsky estimation by orbit determination \citep[see, e.g.][]{milani-gronchi_2009} can evaluate the value of the semi-major axis drift accurately, provided that the observational arc is long enough and the observations are of good quality. These models are typically independent of the physical characteristics of the object. In turn, theoretical models of the Yarkovsky effect rely on the knowledge of the physical properties of the asteroid \citep[see, e.g.][]{bottke-etal_2006}. If the Yarkovsky effect is independently determined from astrometry, the theoretical model of the Yarkovsky effect is constrained, i.e., its solution is known. Therefore, the theoretical model could be inverted, and at least one of its parameters could be estimated.

Yarkovsky effect models depend on several parameters, which are usually the semi-major axis $a$ and the eccentricity $e$ of the asteroid orbit, the diameter $D$, the density $\rho$, the thermal conductivity $K$, the heat capacity $C$, the obliquity $\gamma$, the rotation period $P$, the absorption coefficient $\alpha$, and the emissivity $\varepsilon$. If all the mentioned parameters but the thermal conductivity $K$ are fixed, then the model vs. observed Yarkovsky drift equation 
\begin{equation}
   \left(\frac{\textrm{d}a}{\textrm{d}t}\right) (a, e, D, \rho, K, C, \gamma, P, \alpha,
   \varepsilon) = \bigg(\frac{\textrm{d}a}{\textrm{d}t}\bigg)_\textrm{m},
    \label{eq:yarkoInvertFormula}
\end{equation}
can be solved for $K$, thus determining the thermal conductivity value that allows the model $\textrm{d}a/\textrm{d}t(\, \cdot \,)$ on the left-hand side of Eq.~\eqref{eq:yarkoInvertFormula} to reproduce the observed semi-major axis drift. The choice to determine the parameter $K$ is dictated by the fact that it is the most uncertain one, contrary to the other parameters that can be either estimated or modeled with a fair degree of accuracy. In fact, the thermal conductivity $K$ strongly depends on the characteristics of the material present on the surface of the asteroid, and it can range from $\sim 10^{-5}$ W m$^{-1}$ K${^{-1}}$ for very porous regolith material \citep{2011Icar..214..286K, 2017AIPA....7a5310S}, up to $\sim 50$ W m$^{-1}$ K${^{-1}}$ for very conductive materials such as irons \citep[][and references therein]{2022M&PS...57.1706N}. The Monte Carlo (MC) method presented here assumes that all the parameters but $K$ have some characteristic distribution, and then Eq.~\eqref{eq:yarkoInvertFormula} is solved for $K$ over a large set of combinations of input parameters. These parameters are randomly generated according to the assumed distributions. Solutions of Eq.~\eqref{eq:yarkoInvertFormula} are searched for in a user-defined interval $[K_{\min}, K_{\max}]$, and they are computed with a bisection method with a tolerance of $10^{-11}$ W m$^{-1}$ K${^{-1}}$. Moreover, usually, there is more than one $K$ solution of Eq.~\eqref{eq:yarkoInvertFormula} for a fixed combination of the remaining parameters. By collecting all the solutions obtained for given input combinations, the MC method produces a probability distribution function (PDF) of $K$ in the output. A PDF for thermal inertia $\Gamma$ is then obtained by converting each $K$ solution using the relation
\begin{equation}
    \Gamma = \sqrt{\rho K C},
    \label{eq:thermalInertia}
\end{equation}
and the corresponding values of $\rho$ and $C$. 

The level of accuracy of the estimation of $\Gamma$ that the MC method described above can achieve depends, therefore, on two factors: 1) the accuracy of the Yarkovsky effect modeling, and 2) the reliability of the distributions assumed for the input parameters. In our software, we implemented different Yarkovsky effect models that are suitable for different dynamical and physical scenarios. The details of these models are described in Sec.~\ref{s:yarkoModels}. 

The distributions considered for the orbital and physical parameters also depend on what properties of the studied asteroid are known. Because our method assumes the availability of a measurement of the Yarkovsky effect by astrometry, it implies that the orbital parameters are all determined with very good accuracy, with uncertainties in the semi-major axis typically of the order of $\sim 10^{-9}$ au and of $\sim 10^{-8}$ in eccentricity. Therefore, the orbital parameters are always assumed to be fixed at their nominal values, since their uncertainties produce negligible variations on the predicted Yarkovsky drift. On the other hand, physical properties may be estimated or constrained utilizing different kinds of observations. However, the number of NEAs for which physical properties are determined is still small, and models relying on the general properties of the whole NEA population are needed. In Sec.~\ref{s:physParam}, we describe the models adopted for the input physical parameters.

\section{Yarkovsky effect models}
\label{s:yarkoModels}
The code provides the user with three different models for the computation of the Yarkovsky effect: 1) an analytical model for circular orbits, 2) a semi-analytical model for eccentric orbits, and 3) a semi-analytical model for 2-layered bodies. In addition, variable thermal inertia along the orbit is also implemented in models 2) and 3). We describe hereafter the details of these models. 

\subsection{Analytical model for circular orbits}
\label{s:analyticModel}
The analytical model for the Yarkovsky effect was described in \citet{vokrouhlicky_1998, vokrouhlicky_1999}. The model assumed the asteroid to be a spherical body moving on a circular orbit around the Sun, that rotates around a fixed axis. The boundary conditions for the heat diffusion equation are linearized, and the drift in semi-major axis $(\text{d}a/\text{d}t)$ due to the Yarkovsky effect is given by two distinct components: the seasonal effect $(\text{d}a/\text{d}t)_{\textrm{s}}$, and the diurnal effect $(\text{d}a/\text{d}t)_{\textrm{d}}$. These effects have the following form:
\begin{equation}
    \bigg( \frac{\text{d}a}{\text{d}t} \bigg)_{\textrm{s}} =\hphantom{-} \frac{4 \alpha}{9}
   \frac{\Phi}{\omega_{\textrm{rev}}} F(R'_{\textrm{s}}, \Theta_{\textrm{s}})\sin^2\gamma, 
   \label{eq:yarkoSeasonal}
\end{equation}
\begin{equation}
    \bigg( \frac{\text{d}a}{\text{d}t} \bigg)_{\textrm{d}} = -\frac{8 \alpha}{9}
   \frac{\Phi}{\omega_{\textrm{rev}}} F(R'_{\textrm{d}}, \Theta_{\textrm{d}})\cos\gamma. 
      \label{eq:yarkoDiurnal}
\end{equation}
In the above equations, $\alpha$ is the surface absorption coefficient, $\Phi$ is
the radiation pressure coefficient \citep[e.g.,][]{vokrouhlicky-etal_2015}, $\gamma$ is the spin axis obliquity, and $\omega_{\textrm{rev}}$ is the orbital frequency. 
$R'_{\textrm{s}}$ and $R'_{\textrm{d}}$ are the scaled (non-dimensional) values of the radius $R$, defined as
\begin{equation}
  R'_{\textrm{s}}= \frac{R}{l_{\textrm{s}}}, \quad R'_{\textrm{d}} = \frac{R}{l_{\textrm{d}}},
   \label{eq:rescaledRadius}
\end{equation}
where $l_{\textrm{s}}, \, l_{\textrm{d}}$ are the penetration depths of the seasonal and diurnal thermal waves given by 
\begin{equation}
   l_{\textrm{s}} = \sqrt{\frac{K}{\rho C\omega_{\textrm{rev}}}}, \quad
   l_{\textrm{d}} = \sqrt{\frac{K}{\rho C\omega_{\textrm{rot}}}}.
   \label{eq:penetrationDepth}
\end{equation}
The penetration depths $l_{\textrm{s}}$ and $l_{\textrm{d}}$ depend on the thermal conductivity $K$, the heat capacity $C$, and the density $\rho$ of the asteroid. Additionally, the two length scales depend on the respective frequencies: (i) the spin frequency $\omega_{\textrm{rot}}=2\pi/P$ in the case of the diurnal effect where $P$ is the rotation period, and (ii) the orbital frequency $\omega_{\textrm{rev}}$ in the case of the seasonal effect.
The thermal parameters $\Theta_{\textrm{s}}, \Theta_{\textrm{d}}$ also depend on the physical and thermal characteristics
of the object, and they are defined as 
\begin{equation}
   \Theta_{\textrm{s}} = \frac{\sqrt{\rho K C \omega_{\textrm{rev}}}}{\varepsilon \sigma T_\star^3}, 
    \quad
   \Theta_{\textrm{d}} = \frac{\sqrt{\rho K C \omega_{\textrm{rot}}}}{\varepsilon \sigma T_\star^3}.
    \label{eq:thd}
\end{equation}
In the above equations, $\sigma$ is the Stefan-Boltzmann constant, $\varepsilon$ is the emissivity and
$T_\star$ is the subsolar temperature, defined by $\varepsilon\sigma T_\star^4 = \alpha
\mathcal{E}_\star$, with $\mathcal{E}_\star$ being the solar radiation flux at the
distance of the body. 

\begin{figure}
    \centering
    \includegraphics[width=0.48\textwidth]{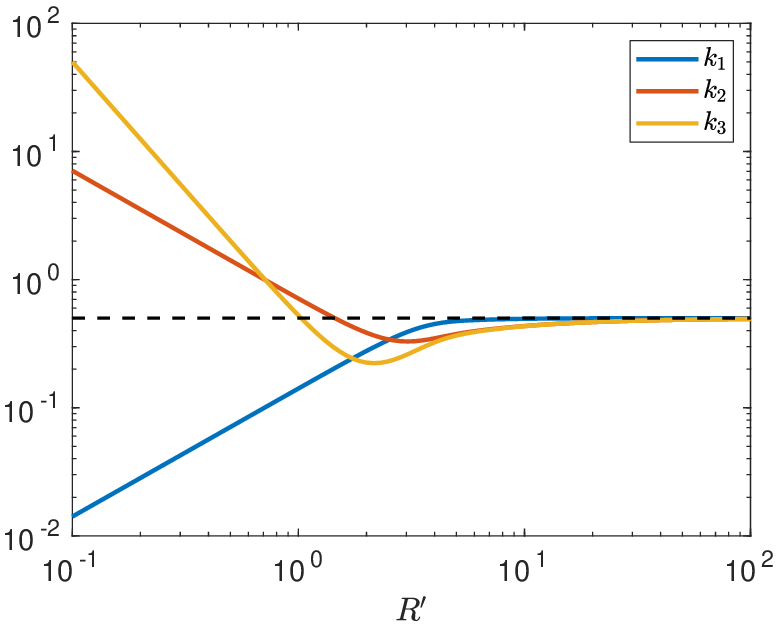}
    \caption{Coefficients $k_1, k_2$, and $k_3$ as a function of the scaled radius $R'$.}
    \label{fig:k1k2k3}
\end{figure}

The function $F$ in Eqs.~(\ref{eq:yarkoSeasonal}) and (\ref{eq:yarkoDiurnal}) is given by
\begin{equation}
    F(R', \Theta) = - \frac{k_1(R')\, \Theta}{1+2k_2(R')\,\Theta + k_3(R')\,\Theta^2}.
    \label{eq:Fs}
\end{equation}
It depends on both the corresponding scaled radius and the thermal parameter. The coefficients $k_1, k_2, k_3$ are positive analytical functions of the scaled radius, and their complete definition can be found in \citet{vokrouhlicky_1998,vokrouhlicky_1999}. Figure~\ref{fig:k1k2k3} shows these coefficients as a function of $R'$. For $R'$ large enough, all the coefficients approach the value $1/2$. It is worth noting that the seasonal component always produces an inward migration, while the direction of migration for the diurnal component depends on the obliquity $\gamma$, which is maximized for $\gamma=0^\circ, 180^\circ$.

It is also important to note here that a double-peaked distribution is often seen in the thermal inertia estimates (see, e.g., Fig.~\ref{fig:bennu_ti_dis}), and it is a consequence of Eq.~\eqref{eq:Fs}. In most cases, one of the peaks could be discarded either based on the number of possible solutions or rejected due to unphysical values of thermal inertia (e.g., $\Gamma > 2000$ J m$^{-2}$ K$^{-1}$ s$^{-1/2}$).

\subsection{Semi-analytical model}
\label{s:semianalyticalModel}
When an asteroid's orbital eccentricity is large, the model described in Sec.~\ref{s:analyticModel} may not be accurate for estimating the Yarkovsky effect. In this case, we can compute the instantaneous semi-major axis drift caused by the Yarkovsky effect as
\begin{equation}
    \bigg(\frac{\textrm{d}a}{\textrm{d}t}\bigg)_{\textrm{i}} = \frac{2}{n^2 a} \mathbf{f}_{\textrm{Y}} \cdot \mathbf{v},
    \label{eq:dadt_i}
\end{equation}
where $a$ is the semi-major axis of the orbit, $n$ is the mean motion, $\mathbf{v}$ is the heliocentric orbital velocity, and $\mathbf{f}_{\textrm{Y}}$ is the instantaneous value of the Yarkovsky acceleration.
The term $\mathbf{f}_{\textrm{Y}} $ is computed with analytical formulas by \citet{vokrouhlicky-etal_2017}, again assuming a homogeneous spherical shape of the asteroid and a linearization of the surface boundary condition. This acceleration is given by
\begin{equation}
     \mathbf{f}_{\textrm{Y}} = \mathbf{f}_{\textrm{Y, d}} + \mathbf{f}_{\textrm{Y, s}},
     \label{eq:yarko_add}
\end{equation}
where $\mathbf{f}_{\textrm{Y, d}}, \mathbf{f}_{\textrm{Y, s}}$ are the diurnal and the seasonal component, respectively.
The diurnal component is expressed as
\begin{equation}
    \mathbf{f}_{\textrm{Y, d}} = \kappa [(\mathbf{n} \cdot \mathbf{s})\mathbf{s} + \gamma_1 (\mathbf{n} \times \mathbf{s}) + \gamma_2 \mathbf{s} \times (\mathbf{n} \times \mathbf{s})].
    \label{eq:f_Yd}
\end{equation}
In Eq.~\eqref{eq:f_Yd}, $\mathbf{n} = \mathbf{r}/r$ is the heliocentric unit position vector, and $\mathbf{s}$ is the unit vector of the asteroid spin axis. In addition,
\begin{equation}
    \kappa = \frac{4 \alpha}{9} \frac{S \mathcal{E}_\star}{m c},
\end{equation}
where $S = \pi R^2$ is the cross-section of the asteroid, $\mathcal{E}_\star$ is the solar radiation flux at a heliocentric distance $r$, $m$ is the asteroid mass, and $c$ is the speed of light. The coefficients $\gamma_1, \gamma_2$ are expressed as
\begin{equation}
    \begin{split}
        \gamma_1(R'_{\textrm{d}}, \Theta_{\textrm{d}}) & = -\frac{k_1(R'_{\textrm{d}}) \, \Theta_{\textrm{d}}}{1 + 2 k_2(R'_{\textrm{d}}) \, \Theta_{\textrm{d}} + k_3(R'_{\textrm{d}}) \,\Theta_{\textrm{d}}^2},\\
        \gamma_2(R'_{\textrm{d}}, \Theta_{\textrm{d}}) & = -\frac{1 + k_2(R'_{\textrm{d}}) \,\Theta_{\textrm{d}}}{1 + 2k_2(R'_{\textrm{d}}) \,\Theta_{\textrm{d}} + k_3(R'_{\textrm{d}})\, \Theta_{\textrm{d}}^2}, 
    \end{split}
    \label{eq:gamma1gamma2}
\end{equation}
where $\Theta_{\textrm{d}}$ is defined by Eq.~\eqref{eq:thd} and $k_1,k_2,k_3$ are the same coefficients as in Sec.~\ref{s:analyticModel}.
The seasonal component is given by
\begin{equation}
    \mathbf{f}_{\textrm{Y, s}} = \kappa [ \bar{\gamma}_1 (\mathbf{n} \cdot \mathbf{s}) + \bar{\gamma}_2 (\mathbf{N}\times \mathbf{n})\cdot \mathbf{s} ]\mathbf{s},  
\end{equation}
where $\mathbf{N}$ is the unit vector normal to the orbital plane, and $\bar{\gamma}_1, \bar{\gamma}_2$ have the same expressions as Eq.~\eqref{eq:gamma1gamma2}, but evaluated with the scaled radius $R'_{\textrm{s}}$ of Eq.~\eqref{eq:rescaledRadius} and thermal parameter $\Theta_{\textrm{s}}$ of Eq.~\eqref{eq:thd}.

The average Yarkovsky drift $\textrm{d}a/\textrm{d}t$ is then obtained by averaging the instantaneous Yarkovsky drift of Eq.~\eqref{eq:dadt_i} over an orbital period as
\begin{equation}
    \frac{\textrm{d}a}{\textrm{d}t} = \frac{1}{2\pi}\int_0^{2\pi} \bigg(\frac{\textrm{d}a}{\textrm{d}t}\bigg)_{\textrm{i}} \, \textrm{d}\ell,
    \label{eq:dadt_semianalytical}
\end{equation}
where $\ell$ is the mean anomaly. The integral at the right-hand side of Eq.~\eqref{eq:dadt_semianalytical} is computed with the trapezoid rule \citep[see, e.g.][]{bulirsch:02}. To prevent numerical errors at high eccentricity, we computed the integral of Eq.~\eqref{eq:dadt_semianalytical} by using a fixed step in the eccentric anomaly $u$, which provides a better sampling of the orbit near the perihelion.

\subsection{Semi-analytical 2-layer model}
\label{s:yarko2L}
In-situ observations made by spacecrafts in the last two decades showed that asteroids have many different surface structures. Some of them, like Bennu or Ryugu, show very rough surfaces composed of boulders spanning orders of magnitude in diameter, basically without any area covered by fine regolith \citep{lauretta-etal_2019, watanabe-etal_2019, sugita-etal_2020}. Other objects like Itokawa or Eros present large smooth areas covered with fine dust and regolith \citep{cheng_2002, miyamoto-etal_2007}, similar to the surface of the Moon. At the same time, their interior may be constituted by much larger boulders, thus breaking the homogeneity assumption made in the Yarkovsky models described in Sec.~\ref{s:analyticModel} and \ref{s:semianalyticalModel}. This last case is particularly important for correctly estimating the Yarkovsky effect because regolith layers are typically composed of loose agglomerates of small rocky particles with large porosity, decreasing the thermal conductivity $K$ with respect to typical values of bare rocks. Since the thermal parameter $K$ can largely change the magnitude of the Yarkovsky effect, the 2-layer model is essential to model regolith-covered asteroids correctly. 

In \citet{vokrouhlicky-broz_1999}, the authors considered a spherical body composed of a core with parameters $\rho_2, C_2, K_2$, covered by a shell of thickness $h$ with parameters $\rho_1, C_1, K_1$. In order to find a solution to the heat conductivity equation, linearized boundary conditions on the outer shell were considered. In addition, a continuity condition on the temperature distribution $T$ and the heat flux when passing between the layers was also assumed. Under these assumptions, the authors found an analytical solution for the seasonal Yarkovsky effect and discussed the difficulties in finding a solution for the diurnal effect for meter-sized objects. The problem of finding analytical solutions to the heat diffusion equation on non-homogeneous bodies was also discussed in \citet{capek-vokrouhlicky_2012}. A complete analytical model for the Yarkovsky effect on 2-layered asteroids is unavailable. 

Due to the above difficulties, we used a simplified approach to define a 2-layer Yarkovsky model. We assumed the surface material to have a density of $\rho_{\textrm{surf}}$, while the interior to have a density $\rho$. To define the thickness of the top layer, we assumed that the penetration depths of the seasonal and diurnal thermal waves are smaller than the real regolith thickness. At least roughly speaking, in this case, the magnitude of the Yarkovsky effect depends only on the heat transport in the top layer. 

The second layer represents the body interior, excluding only the small top layer. It is assumed, however, that the second (interior) layer does not affect the temperature distribution at the surface. It is taken into account only when asteroid mass is computed. 

Therefore, the thickness of the top layer could be approximated with the length of the more relevant penetration depth, which is the one related to the more important component of the Yarkovsky effect. In most cases, this is the diurnal penetration depth $l_\textrm{d}$, except when an asteroid's obliquity is close to 90 degrees when it is assumed that the top layer corresponds to the seasonal penetration depth $l_\textrm{s}$. 

The two penetration depths are obtained as
\begin{equation}
       l_{\textrm{s}} = \sqrt{\frac{K}{\rho_{\textrm{surf}} C\omega_{\textrm{rev}}}}, \quad
   l_{\textrm{d}} = \sqrt{\frac{K}{\rho_{\textrm{surf}} C\omega_{\textrm{rot}}}}.
   \label{eq:penetrationDepth2L}
\end{equation}

Once the top layer has been constrained, we used its adopted parameters to compute the Yarkovsky effect with Eq.~\eqref{eq:dadt_i}. 

We recall that this Yarkovsky model is valid only as long as the penetration depths of Eq.~\eqref{eq:penetrationDepth2L} are smaller than the actual regolith thickness, which is typically satisfied when $R'_{\textrm{s}}$ and $R'_{\textrm{d}}$ are $>> 1$. Therefore, it is the user's responsibility to make use of this model with more caution by checking that the values of $l_{\textrm{s}}$ and $l_{\textrm{d}}$ obtained in output are reasonably smaller than the presumed thickness of the surface material. Generally, the approach used for the 2-layer model is an approximation. In particular, we found that it generally tends to underestimate the dust layer's role. Therefore, in its current form, the 2-layer model may only indicate how significantly a possible presence of the surface dust layer may affect the thermal inertia estimation by the ASTERIA.

\subsection{Variable thermal inertia}
Thermal inertia depends on the thermal conductivity and heat capacity through Eq.~\eqref{eq:thermalInertia}. Laboratory analyses on lunar regolith samples \citep{keihm_1984} and meteorites \citep{2020M&PS...55E...1O} showed that both thermal conductivity $K$, and heat capacity $C$ depend on the temperature. Therefore, the thermal inertia also depends on the temperature, which is, in turn, related to the heliocentric distance $r$. Accordingly, accounting for its variation along the asteroid orbit is important to estimate the thermal inertia correctly. 

For predominantly radiative heat transfer, as expected in a fine regolith on an airless body, the thermal inertia scales with 
\begin{equation}
    \Gamma = \Gamma_0 r^{-3/4},
    \label{eq:TItheoretical}
\end{equation}
where $\Gamma_0$ is the value of the thermal inertia at 1 au.
In a more recent study, \citet{rozitis-etal_2018} found that the trend of Eq.~\eqref{eq:TItheoretical} found by theoretical models does not always agree with the observations, but more extreme values of the exponent may be possible. Similar conclusions were also drawn by \citet{2021PSJ.....2..161M}.

To account for the variable thermal inertia in our model, yet keeping it flexible concerning the exact dependence on the heliocentric distance, we assume that the thermal inertia $\Gamma$ varies with the heliocentric distance as
\begin{equation}
    \Gamma = \Gamma_0 r^{\beta}.
    \label{eq:TIimplemented}
\end{equation}

The exponent $\beta$ is left as a free parameter to be chosen by the user. The variable thermal inertia is implemented for the Yarkovsky models described in Sec.~\ref{s:semianalyticalModel} and \ref{s:yarko2L}. On the purely technical side, we underline that when the variable thermal inertia model is used, output from our model is thermal inertia $\Gamma_0$, which is a value normalized to $r=1$~au.

\subsection{Non-sphericity effects}
\label{ss:non-sphericity}

The above-described Yarkovsky models assume a spherical shape of the body. However, it is known that asteroids are not perfectly spherical but rather elongated, in some cases even very elongated bodies \citep[see, e.g.][and references therein]{2019AJ....157..164M}. \citet{vokrouhlicky_1998} found that in the general case, the diurnal Yarkovsky component for spheroids and a sphere does not differ more than 30\%. However, in some cases, the difference could be up to a factor of 2 or 3. Such an example is fixed spin axis orientation at high obliquity.

As the Yarkovsky detection is performed from relatively short time intervals (up to a few tens of years), it is reasonable to consider the orientation of the asteroid spin axis fixed in space. 

Assuming a biaxial ellipsoid shape (spheroid), rotation around the shortest axis, and high obliquity, an elongated object has a smaller average cross-section than a spherical body of the same equivalent diameter. This implies that elongated bodies are subject to a smaller Yarkovsky effect than spherical ones of the same equivalent diameter. Therefore, to account for the impact of the body's non-sphericity, we introduced a correction factor $\xi$ into the model. We found that the non-sphericity correction factor could be reasonably well approximated as $\xi = f^{-0.3}$. The factor is defined to mimic the results presented in \citet{vokrouhlicky_1998}. The value $f$ is the ratio of the $a/b$ axes and is computed from the maximum light curve amplitude $\Delta m$ using the relation $\Delta m = 2.5 \log f$. As we assume $a>b>0$, the axis ratio $f$ and the correction factor $\xi$ are always in the $(0,1)$ interval. 

As the non-sphericity under the given assumptions decreases the Yarkovsky effect, and the correction factor expressed the magnitude of the decline, to compensate for the non-sphericity, we divided the measured drift in the semi-major axis $\textrm{d}a/\textrm{d}t$ with the factor $\xi$. In other words, we artificially increased the measured Yarkovsky drift to match the one expected for a spherical object having the same equivalent diameter. We point out to use this scaling with caution as it might break down for extreme light curve amplitudes $\Delta m \gtrsim 2.0$.

\section{Modeling the input physical parameters}
\label{s:physParam}
\subsection{Albedo distribution}
\label{ss:albedo_dis}
The albedo $p_V$ is typically determined from the infrared observations. It is nowadays available for a large number of asteroids thanks to space surveys dedicated to this purpose, such as the IRAS \citep{2002AJ....123.1056T}, the WISE \citep[][]{2011ApJ...741...68M, 2015ApJ...814..117N}, and the AKARI mission \citep{2018A&A...612A..85A}.

When albedo and its associated uncertainty are known, the user can select an option that uses the nominal value and assumes a Gaussian distribution for the measurement error.

If $p_V$ is not known, we model this parameter by combining the NEA absolute magnitude and orbital distribution by \citet{granvik-etal_2018}, and the NEA albedo distribution by \citet{morbidelli-etal_2020}. A first description of this model for the NEA 2011 PT was given in \citet{2021A&A...647A..61F}. Given the orbital elements $(a,e,i)$ of the NEA and its absolute magnitude $H$, the model by \citet{granvik-etal_2018} provides 7 probabilities $P_s(a,e,i,H), s=1,\dots,7$. Each of these numbers corresponds to the probability that the NEA arrived in the NEA region through one of the 7 seven transport routes from the main belt: 1) the $\nu_6$ secular resonance; 2) the 3:1 Jupiter mean-motion resonance (JMM); 3) the 5:2 JMM; 4) the Hungaria region; 5) the Phocaea region; 6) the 2:1 JMM; 7) the Jupiter Family Comets (JFC) region. In the NEA albedo distribution by \citet{morbidelli-etal_2020}, asteroids are divided into three albedo categories: $c_1$ corresponds to dark objects with $p_V \leq 0.1$, $c_2$ corresponds to moderately bright objects with albedo $0.1 < p_V \leq 0.3$, and $c_3$ corresponds to bright objects with $p_V > 0.3$. For each source region, indexed with $s$, the model provides the fraction of objects $p_s(c_i), i=1,2,3$ belonging to the category $c_i$ that arrived from the source region $s$. Numerical values are given in Table~\ref{tab:albedoFractions}. 
\begin{table}[ht]
    \caption{The albedo probability for each NEAs source region, as given in \cite{morbidelli-etal_2020}.}
    \centering
    \begin{tabular}{cccc}
         \hline
         \hline
    Source Region & $p_s(c_1)$ & $p_s(c_2)$ & $p_s(c_3)$ \\
         \hline
    $\nu_6$  & 0.120 & 0.558 & 0.322  \\
    3:1      & 0.144 & 0.782 & 0.074  \\
    5:2      & 0.294 & 0.557 & 0.149  \\
    Hungaria & 0.021 & 0.113 & 0.866  \\
    Phocaea  & 0.501 & 0.452 & 0.047  \\
    2:1      & 0.399 & 0.200 & 0.401  \\
    JFC      & 1.000 & 0.000 & 0.000  \\
         \hline
    \end{tabular}
    \label{tab:albedoFractions}
\end{table}

To define a PDF for the albedo of a specific object, we first define an albedo PDF $p_s(p_V)$ for each escape route $s$. This PDF is uniform in the categories $c_1$ and $c_2$, and exponentially decaying in $c_3$ \citep{morbidelli-etal_2020} as
\begin{equation}
    p_s(p_V) = 
    \begin{cases}
    \displaystyle \frac{p_s(c_1)}{0.1}, & p_V \leq 0.1,    \\[2ex]
    \displaystyle \frac{p_s(c_2)}{0.2}, & 0.1<p_V\leq 0.3, \\[2ex]
    \displaystyle p_s(c_3)\frac{2.6^{-\frac{p_V-0.3}{0.1}}}{\int_{0.3}^1 2.6^{-\frac{x-0.3}{0.1}}dx}, & p_V > 0.3.
    \end{cases}
    \label{eq:pspv}
\end{equation}
The albedo PDF $p(p_V)$ is then defined as
\begin{equation}
    p(p_V) = \sum_{s=1}^7 P_s(a,e,i,H)\, p_s(p_V).
    \label{eq:PDF_albedo}
\end{equation}

\subsection{Diameter and density distributions}
The input diameter and density distributions could be either population-based or user-provided. The diameter of an asteroid can be determined with different methods, including radar observations \citep{1985PASP...97..877O}, thermal infrared modeling \citep{1998Icar..131..291H}, and stellar occultation \citep{1979aste.book...98M}. If the asteroid's diameter is measured, the user can specify the mean value and the 1-$\sigma$ uncertainty to model $D$ with a Gaussian distribution.

On the other hand, if the diameter is not available, it could be estimated by the conversion formula \citep[see, e.g.][]{bowell-etal_1989, 2007Icar..190..250P} 
\begin{equation}
   D = \frac{1329 \textrm{ km}}{\sqrt{p_V}}10^{-H/5}.
   \label{eq:mag2dia}
\end{equation}

The density is even more challenging to measure. Typical methods to determine the mass of an asteroid are through close encounters with smaller bodies, spacecraft tracking, and the orbit of a satellite \citep{carry_2012}. Then, the density $\rho$ is obtained by combining the estimate for $m$ and that of the diameter $D$. Therefore, direct density determination is available for a small number of asteroids. Nevertheless, to keep the code flexible, we allow the user to choose a normal distribution for the density $\rho$, with user-specified mean and standard deviation values. 

Another indirect way to have an estimate of the density is through the spectroscopic classification \citep{1984PhDT.........3T, demeo-etal_2009, 2010A&A...510A..43C}. Indeed, asteroids with different spectra are supposed to have different compositions. The density of each taxonomic class can be assumed as the same of other asteroids with the same spectral class for which the density was determined with one of the methods mentioned above \citep[see][for a review about asteroid densities]{carry_2012}, or by the association of the spectral type with known meteorites \citep[see, e.g.][]{binzel-etal_2019,2022Icar..38014971D}. 

When estimates of $D$ and $\rho$ are unavailable, we model these parameters with population-based properties. The model also relies on the fact that diameter $D$ and the density $\rho$ are not uncorrelated because objects with large albedo correspond to smaller asteroids with relatively large density, and objects with low albedo correspond to larger asteroids with relatively low density. Therefore, we generate a joined $(D, \rho)$ distribution using geometric albedo $p_V$ obtained as explained in Section~\ref{ss:albedo_dis}. A sample of $p_V$ is obtained first, and then each sample value is converted into $D$ and $\rho$. The diameter $D$ is obtained through Eq.~\eqref{eq:mag2dia}, where $H$ is assumed to be Gaussian distributed. The density $\rho$ is instead generated according to the category $c_j$ into which $p_V$ falls. We associate category $c_1$ to the C-complex of carbonaceous asteroids, category $c_2$ to stony asteroids in the S-complex, and category $c_3$ to bright asteroids of the X-complex. Then, we randomly generate a value of $\rho$ depending on the complex associated with the value of $p_V$. The densities of the complexes are assumed to be distributed according to a log-normal distribution, with parameters listed in Table~\ref{tab:astDensities}. 

\begin{table}[!ht]
    \caption{Parameters of the log-normal distributions of the density for the three different complexes of asteroids.}
    \centering
    \begin{tabular}{ccc}
    \hline
    \hline
         Complex & $\rho$ (kg m$^{-3}$) & $\sigma_{\rho}$ (kg m$^{-3}$) \\
         \hline
            C    & 1200  & 300 \\
            S    & 2720  & 540 \\
            X    & 2350  & 520 \\
         \hline
    \end{tabular}
    \label{tab:astDensities}
\end{table}

\subsection{Surface density distribution}
The two-layer model of Sec.~\ref{s:yarko2L} also needs the distribution of the density of the surface material $\rho_{\textrm{surf}}$ in input. As discussed above, this model is supposed to be used only in the case of a body covered by a regolith layer. Even though regolith comprises a large variety of dust and gravel sizes, it generally refers to an unconsolidated, loose, porous, and heterogeneous set of small rocky materials. 

Thanks to in-situ observations by spacecrafts and to the samples returned from the Moon surface during the Apollo missions, the regolith density has been constrained by direct measurements \citep{1972LPSC....3.3235M}. The average density of the lunar surface regolith is about 1300 kg~m$^{-3}$. Recent works based on the lunar far side obtained values for the upper top level as low as about 500 kg~m$^{-3}$ \citep{10.1093/nsr/nwac175}. We allow the user to use only a Gaussian distribution for $\rho_{\textrm{surf}}$, by specifying the mean and standard deviation values. The critical point is that the two-layer model makes sense only if the surface density is significantly lower than the bulk density. Therefore, values above 1300 kg~m$^{-3}$ should be avoided. For example, we consider the values of 1100$\pm$100 kg~m$^{-3}$, as a good compromise. Nevertheless, the users are welcome to test other values. 

\subsection{Obliquity distribution}
The obliquity of an asteroid is typically determined by light curve inversion method that makes use of photometric observations \citep{kaasalainen-torppa_2001, kaasalainen-etal_2001, kaasalainen-etal_2002, durech-etal_2015}, or by radar observations \citep{hudson_1994, hudson-etal_2000, magri-etal_2007}. When determinations are available, the user can decide to use a Gaussian distribution with a mean value equal to the nominal estimated obliquity, and standard deviation equal to the 1-$\sigma$ uncertainty. 

When the obliquity is unknown, we rely again on the global properties of the NEAs population, and we model this parameter with the NEA obliquity distribution found by \citet{tardioli-etal_2017}. The distribution of $\gamma$ is best fitted by a quadratic PDF in $\cos\gamma$ of the form
\begin{equation}
    p(\cos\gamma) = a \cos^2\gamma + b \cos\gamma + c,
    \label{eq:PDF_gamma}
\end{equation}
where $a=1.12, b=-0.32$, and $c = 0.13$. 

\begin{figure}[!ht]
    \centering
    \includegraphics[width=0.48\textwidth]{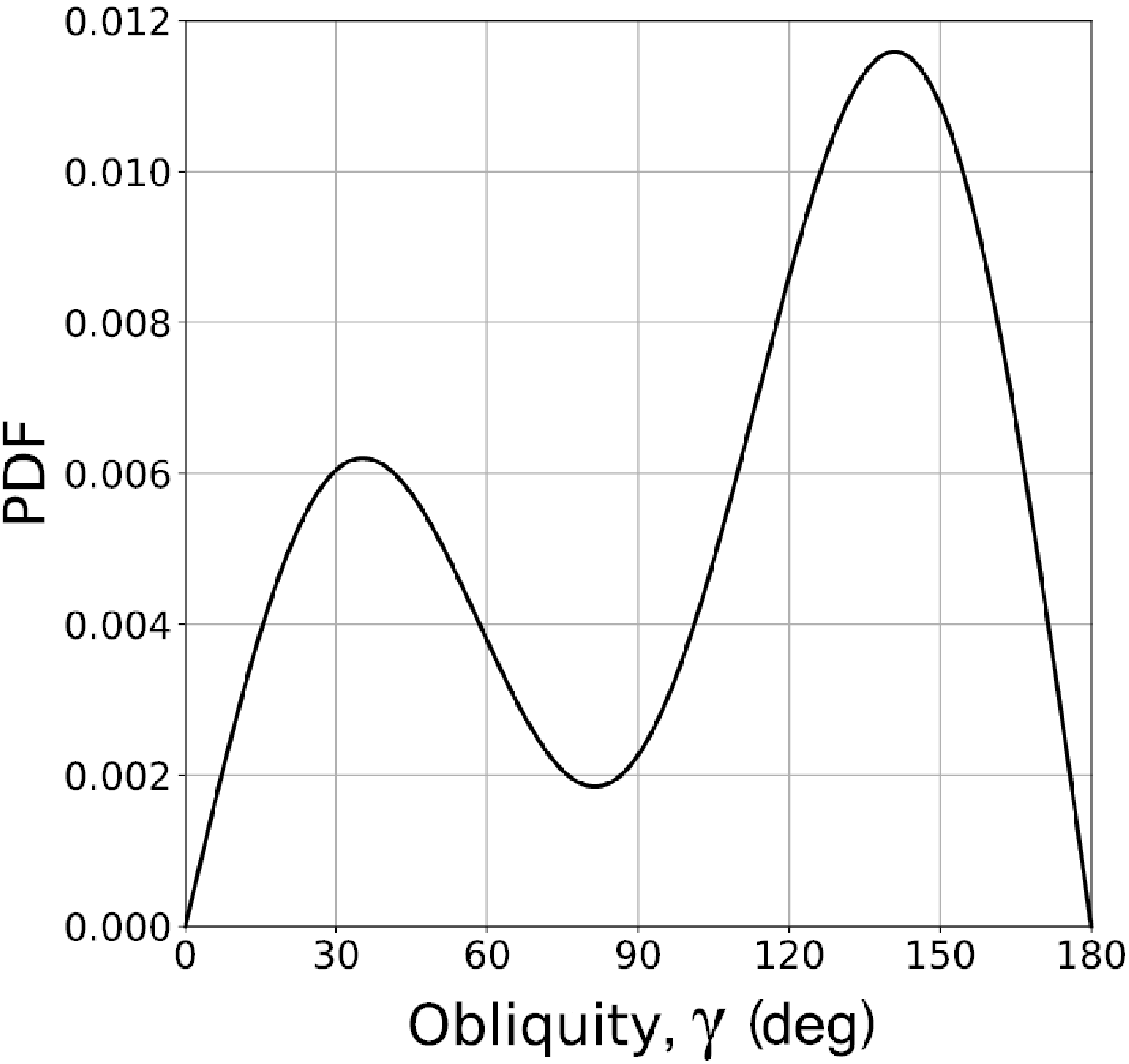}
    \caption{PDF of the obliquity distribution of the NEAs population.}
    \label{fig:pdf_cosgamma}
\end{figure}
The resulting probability density function of the NEAs' obliquities derived from Eq.~\eqref{eq:PDF_gamma} is shown in Fig.~\ref{fig:pdf_cosgamma}. The adopted distribution has about a 2:1 ratio between retrograde and prograde rotators \citep{2004Natur.428..400L} and represents the recent obliquity model of near-Earth objects population \citep{tardioli-etal_2017}. In practice, this should not affect the final result produced by the ASTERIA model. The values of $\gamma$ not compatible with the measured Yarkovsky drift are always rejected.

\subsection{Rotation period distribution}
The rotation period $P$ is an essential parameter for correctly estimating the Yarkovsky effect because it may change the magnitude by a factor of several. It is typically evaluated by analyzing the light curve, and most of the measurements can be found in the online Asteroid Light curve Database\footnote{\url{https://alcdef.org/}} \citep[LCDB;][]{warner-etal_2009}. 

While the rotation period of NEAs generally shows a trend with the diameter $D$, modeling $P$ in different size categories based on NEA population properties would produce poor results in estimating the thermal inertia $\Gamma$ with the MC method. This is because the distribution of $P$ has a large standard deviation, meaning that a population-based distribution is not very representative of the true rotational state the asteroid is in. Additonally, \citet{2020ApJS..247...26P} found that the number of bodies with long rotation periods is underestimated by the ground-based surveys, implying that the LCDB catalogue could be biased. For these reasons, we decided not to model $P$ with population-based information, and we limited ourselves to providing the possibility of using a Gaussian distribution model with user-specified parameters. Therefore, the period needs to be known, or the model should be run for at least several discrete values.

\subsection{Heat capacity}
The heat capacity value $C$ depends on the physical characteristics of the materials composing the asteroid. It can vary within a factor of several \citep{delbo-etal_2015}. Typical values assumed for rocky and regolith-covered main-belt asteroids range from 600 to 750 J kg$^{-1}$ K$^{-1}$, while they go as low as 500 J kg$^{-1}$ K$^{-1}$ for iron-rich asteroids \citep{farinella-etal_1998}. In addition, the heat capacity shows to increase with increasing temperatures, and therefore, it could be higher for NEAs, though not exceeding $\sim$1200 J kg$^{-1}$ K$^{-1}$ \citep[e.g.][]{2020M&PS...55E...1O}. However, there is insufficient data to extrapolate a general distribution valid for the whole NEA population. Therefore, the model keeps the heat capacity $C$ fixed, and the user can specify its value. We anyway suggest keeping $C$ within the reasonable ranges discussed above. We also note that the thermal inertia estimations by the ASTERIA model are not sensitive to changes in $C$. However, the code also provides an opportunity to estimate the thermal conductivity. In the latter case, caution is needed when selecting the input heat capacity, and we suggest running the code with several different values.

\subsection{Absorption coefficient and emissivity}
The absorption coefficient is defined as $\alpha = 1 - A$, where $A$ is the Bond albedo. The Bond albedo $A$ is related to the geometric albedo $p_V$ by 
\begin{equation}
    A = q p_V.
    \label{eq:bondAlbedo}
\end{equation}
In the above formula, $q$ is the phase integral defined by
\begin{equation}
    q = 0.29 + 0.684 G,
\end{equation}
where $G$ is the slope parameter. If the user provides a measurement of the geometric albedo $p_V$, Eq.~\eqref{eq:bondAlbedo} is used to produce a distribution of $\alpha$, which therefore follows Gaussian distribution. On the other hand, the population-based albedo distribution of Eq.~\eqref{eq:PDF_albedo} is used to produce a distribution of $\alpha$. In all the computations, the phase angle is assumed to be $G=0.15$.

The emissivity $\varepsilon$ can be constrained by measurements performed on meteorite samples. \citet{ostrowsky-bryson_2019} collected the measurements of $\varepsilon$ of 61 meteorites, and all the objects but one have a value ranging between 0.9 and 1, with an average value of 0.984. Since the interval in which the emissivity may change is small, we expect changes in the estimated Yarkovsky effect to be small, and therefore we decided to keep this parameter fixed in the MC method. The value can be specified by the user, but if there is no other reason to believe it to be smaller than 0.9, we suggest using the average value $\varepsilon = 0.984$ found for meteorite samples. 

\subsection{Measured Yarkovsky effect distribution}
As mentioned in Sec.~\ref{s:MC}, the availability of a measurement of the Yarkovsky effect obtained by astrometry is a fundamental requirement for our MC method. Therefore, we always assume the semi-major axis drift $(\textrm{d}a/\textrm{d}t)_{\textrm{m}}$ to be Gaussian distributed with a mean value corresponding to the nominal value, and standard deviation equal to the 1-$\sigma$ uncertainty. These values can be specified by the user in the input of the software.

\section{ASTERIA software}
\label{s:software}
\subsection{Open-source software and repository}
All the methods and models described in Sec.~\ref{s:MC}, \ref{s:yarkoModels}, and \ref{s:physParam} are implemented in the ASTERIA software v1.0.0 \citep{asteria_zenodo}, which is written in \texttt{fortan90} programming language. A publicly available repository is also maintained on GitHub\footnote{\url{https://github.com/Fenu24/D-NEAs}}. The ASTERIA software is released under the Attribution-NonCommercial-ShareAlike 4.0 International (CC BY-NC-SA 4.0) license. Automatic scripts, working under UNIX systems, for the compilation of the source code with the GNU \texttt{gfortran} or Intel \texttt{ifort} compilers, are included in the distribution. 

Once compiled, the ASTERIA software provides the user with two executable drivers: 1) the \texttt{gen\_distrib.x} driver, implementing the methods described in Sec.~\ref{s:physParam}, and 2) the \texttt{gamma\_est\_mc.x} driver, implementing the MC method described in Sec.~\ref{s:MC}. 
The driver \texttt{gen\_distrib.x} needs to be executed first in order to produce the input distributions of the physical parameters entering the Yarkovsky modeling. The files generated are then used by the \texttt{gamma\_est\_mc.x} driver for the MC method of thermal inertia estimation. This separation was done to keep the code more flexible. The user is strongly encouraged to use the \texttt{gen\_distrib.x} driver for generating the input distributions, but one can also produce their own files with parameters' distributions and use them as input for the \texttt{gamma\_est\_mc.x} driver.
The driver \texttt{gamma\_est\_mc.x} for the MC estimation of thermal inertia has been parallelized by using the \texttt{OpenMP} API, and it is, therefore, able to run on multi-CPUs machines. Detailed instructions for the compilation and for the usage of these drivers can be found in the Software User Manual included in the ASTERIA software repository.

\section{Model validation}
\label{s:validation}

Model validation is a key procedure for quantifying the model's accuracy, i.e., determining how well the model describes the actual conditions. Validation aims to quantify confidence in the model's predictive capability by comparison with available data.

To validate our model, we selected a set of eleven asteroids with thermal inertia estimates available in the literature as determined by means of other models and techniques. The group essentially consists of the asteroid Bennu plus ten other near-Earth asteroids. The asteroid (101955) Bennu, a target of NASA's OSIRIS-REx mission, is the most suitable object for this purpose. It has practically all parameters relevant to the ASTERIA model very well constrained  (Table~\ref{tab:bennu}), including the thermal inertia, which is the output of the model. Additional objects selected for our validation purposes are those NEAs well-characterized by ground- and space-based telescopes. Therefore, we first validated our model on Bennu as the most reliable reference\footnote{We recall here that, being visited by the Hayabusa-2 mission, naturally, asteroid Ryugu is expected to provide the same validation opportunity as Bennu. Unfortunately, due to its larger size, the Yarkovsky detection for Ryugu is either unavailable or poorly constrained, preventing us from using the asteroid for our validation purposes. }, then extended it to ten other NEAs. The validation on Bennu aims to show that our model can reproduce the thermal inertia estimate when the input parameters are well-known. In contrast, the test on the other objects primarily seeks to show that the model gives reliable results in more realistic situations when only some input parameters are known and that the results are consistent with the available estimates in the literature.

\subsection{Model validation on asteroid Bennu}

A preliminary model and ASTERIA code testing on asteroid Bennu have already been performed by \citet{Fenucci_et_al_AA2023}. This work investigated how various unknown input parameters affect the final thermal inertia estimates, and showed that the model can reproduce reasonably well Bennu's thermal inertia. Building on these findings, here we extended this analysis, investigating the other relevant factors that may affect the match between our estimate and the reference value(s). In particular, the factors that were not considered by \citet{Fenucci_et_al_AA2023} are the role of the non-spherical shape and of the spatial variations of Bennu's surface thermal inertia. Referent thermal inertia estimates for asteroid Bennu, used to benchmark our model, are listed in Table~\ref{tab:bennu_TI}. 

\begin{table*}[]
    \caption{Orbital and physical parameters of asteroid (101955) Bennu.}
    \centering
    \begin{tabular}{lcc}
         \hline
         \hline
    Parameter & Value & Reference \\
         \hline
    Semi-major axis, $a$   & 1.12599635679  $\pm$ 1.7$\times 10^{-10}$ au   &  NEOCC - Epoch 60000.0 MJD \\
    Eccentricity, $e$  &  0.203719195 $\pm$ 2.0$\times 10^{-8}$   & NEOCC - Epoch 60000.0 MJD \\
    Semi-major axis drift, $\text{d}a/\text{d}t$  & $-284\pm0.2$ m yr$^{-1}$ & \citet{2021Icar..36914594F} \\
    \hline
    Radius, $r$  & 242.22 $\pm$ 0.15 m & \citet{2020SciA....6.3649D} \\
    Best-fit ellipsoid, $a\times b\times c$  & 252.4$\times$245.9$\times$228.4 m & \citet{2020SciA....6.3649D} \\
    Bulk density, $\rho$  & 1194 $\pm$ 3 kg m$^{-3}$  & \citet{2020SciA....6.3649D} \\
    Obliquity, $\gamma$  & 177.6 $\pm$ 0.11 & \citet{lauretta-etal_2019} \\
    Rotation period, $P$  & 4.2960015 $\pm$ 0.0000018 hours & \citet{2020SciA....6.3649D} \\
    Albedo, $p_V$   & 0.044 $\pm$ 0.002 & \citet{lauretta-etal_2019} \\
    Emissivity, $\epsilon$   & 0.95 &  \citet{2020SciA....6.3699R} \\
         \hline
    \end{tabular}
    \label{tab:bennu}
\end{table*}

For the input parameters, we did not use population-based results but generated corresponding distributions based on the values listed in Table~\ref{tab:bennu}. In addition to the parameters listed in the table, we adopted a heat capacity value\footnote{We recall that the results obtained with our model are not particularly sensitive to the values of heat capacity used \citep{2021A&A...647A..61F, Fenucci_et_al_AA2023}.} of $C=750$ J kg$^{-1}$ K$^{-1}$, following \citet{2020SciA....6.3699R}.

Also, as there are different variations and options in our model, as described in Section~\ref{s:yarkoModels}, we analyzed which one should be the most appropriate for asteroid Bennu, and set up our nominal model settings. In particular, we implemented into the ASTERIA two Yarkovsky models based on circular and eccentric orbit assumptions. In this respect, we underline that the main advantage of the circular model is that it is fast and provides results almost instantaneously. In contrast, the eccentric-orbit-based model is a much more time-consuming procedure. With an eccentricity of about 0.2, the orbit of Bennu is not very eccentric, and even a circular model might be appropriate. However, aiming to obtain as accurate results as possible, we used solely the eccentric Yarkovsky model for Bennu.
Similarly, we decided not to use a two-layer density model, as no fine regolith (dust) has been found at the surface of Bennu \citep{2019NatAs...3..341D}.

\citet{2020SciA....6.3699R} found that thermal inertia variations with temperature (hence with heliocentric distance) at Bennu are small. For that reason, we used constant thermal inertia in our nominal model. 

In addition, we considered other aspects that could affect the results in the case of asteroid Bennu. One of these is the correction due to the non-sphericity of Bennu. As the shape of the Bennu is well-defined, we used the measured axis ratio here to compute the non-sphericity factor, instead of obtaining it from the light curve amplitude.\footnote{We note here that our standard approach described in Section~\ref{ss:non-sphericity} and the maximum light curve amplitude of $\Delta m=0.3$ \citep{warner-etal_2009}, yields a somewhat smaller value of the correction factor of $\xi=0.87$.} The $a/c$ ratio of 1.1 yields the correction factor of $\xi=0.95$.

Furthermore, as discussed by \citet{2021A&A...647A..61F}, nonlinearity effects decrease the theoretically predicted semi-major axis drift rate $\text{d}a/\text{d}t$, with a reduction factor ranging between 0.7 and 0.9. Conversely, thermal beaming effects always increase the semi-major axis drift rate by a factor ranging between 1.1 and 1.5. These two effects tend to compensate for each other, although some residual factors may remain in particular cases. Although we cannot properly resolve these two issues, we attempt to reduce their importance by artificially increasing the uncertainty of the measured semi-major axis drift rate. In this case, we assumed that the uncertainty was 5\% of the nominal value, reducing the corresponding detection level to 20-$\sigma$, significantly lower than the 1400-$\sigma$ obtained by \citet{2021Icar..36914594F}.

Finally, for the emissivity, we used a value of 0.95, as in \citet{2020SciA....6.3699R}. With this in mind, we defined the above-described model settings as nominal. 

\begin{table*}[]
    \caption{Thermal inertia estimations for asteroid (101955) Bennu.}
    \centering
    \begin{tabular}{lcc}
         \hline
         \hline
     Value  & Note & Reference \\
         \hline
    $\Gamma$ = 310$\pm$70 J m$^{-2}$ K$^{-1}$ s$^{-1/2}$ &  & \citet{2014Icar..234...17E} \\
    $\Gamma$ = 350$\pm$20 J m$^{-2}$ K$^{-1}$ s$^{-1/2}$ &  &  \citet{2019NatAs...3..341D}  \\
    $\Gamma$ = 300$\pm$30 J m$^{-2}$ K$^{-1}$ s$^{-1/2}$ & OTES & \citet{2020SciA....6.3699R}  \\
    $\Gamma$ = 320$\pm$30 J m$^{-2}$ K$^{-1}$ s$^{-1/2}$ & OVIRS & \citet{2020SciA....6.3699R}  \\
    $\Gamma$ = 350$\pm$30 J m$^{-2}$ K$^{-1}$ s$^{-1/2}$& OVIRS: equatorial region & estimate based on Fig.~5 from \citet{2020SciA....6.3699R}  \\
%    \hline
    $\Gamma$ = 381$_{-56}^{+55}$ J m$^{-2}$ K$^{-1}$ s$^{-1/2}$ & Yarkovsky effect & this work (nominal model)  \\
    $\Gamma$ = 374$_{-53}^{+54}$ J m$^{-2}$ K$^{-1}$ s$^{-1/2}$ & Yarkovsky effect & this work considering spatial variations in thermal inertia  \\
         \hline
    \end{tabular}
    \label{tab:bennu_TI}
    \tablenotetext{ OSIRIS-REx Thermal Emission Spectrometer (OTES), \\ 
    OSIRIS-REx Visible and InfraRed Spectrometer (OVIRS).}
\end{table*}

Running the model with the nominal settings, the resulting median value of the thermal inertia and its corresponding 1-$\sigma$ lower and upper uncertainties for the higher peak are $\Gamma$=381$_{-56}^{+55}$~J m$^{-2}$ K$^{-1}$ s$^{-1/2}$. The obtained value is generally compatible with all the literature estimates listed in Table~\ref{tab:bennu_TI}.

\begin{figure}[!ht]
    \centering
    \includegraphics[width=0.48\textwidth]{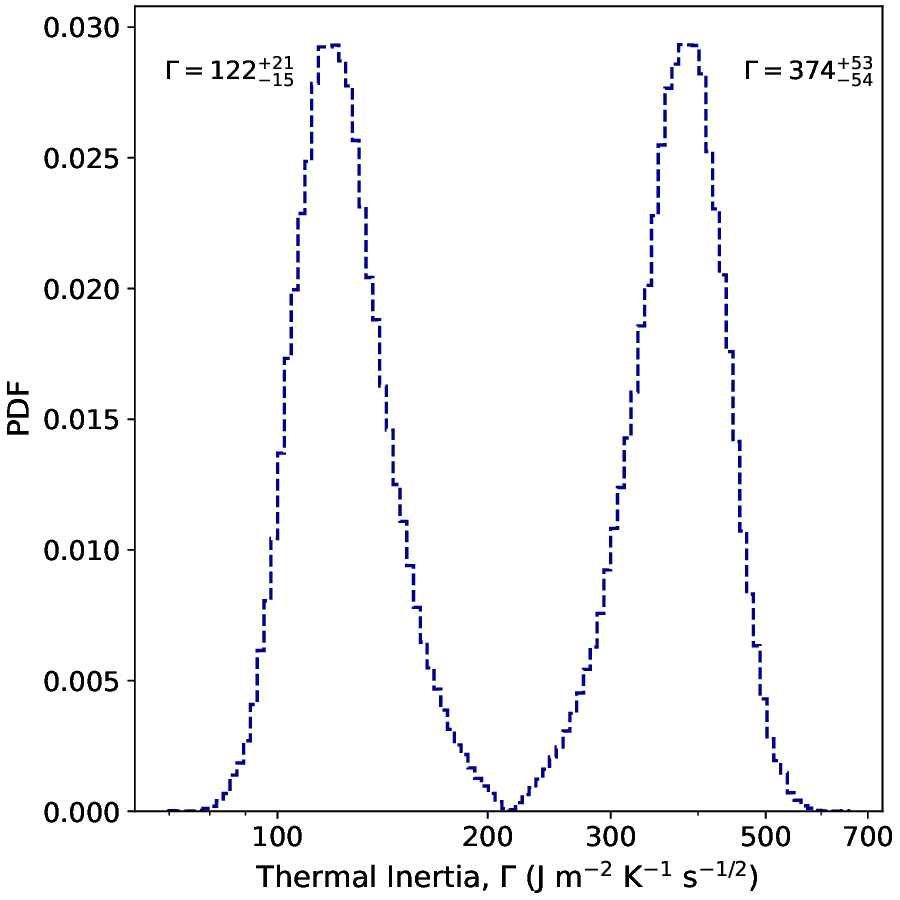}
    \caption{Distribution of asteroid Bennu's thermal inertia $\Gamma$ generated with the ASTERIA model. The figure shows the results for the nominal settings after considering the spatial variations in Bennu's surface thermal inertia. See text for additional details. The median values of $\Gamma$ and their corresponding lower and upper $1-\sigma$ uncertainties are shown for each of the peaks.}
    \label{fig:bennu_ti_dis}
\end{figure}

Still, we note that our value is shifted towards higher values with respect to other estimates.
\textit{What could be the reasons for that?} The ASTERIA model uses the orbit-averaged Yarkovsky effect to, in principle,
estimate the global average thermal inertia. However, the model might, to some degree, depend on the
spatial variations of the surface thermal inertia. As such variations have been observed on Bennu \citep{2020SciA....6.3699R}, we investigate how they could affect the results. 

\citeauthor{2020SciA....6.3699R} found larger thermal inertia at Bennu's equator than at polar regions.
Assuming spatial variations in thermal inertia, the quantity relevant to Yarkovsky is proportional to
$\Gamma(\varphi) cos^{2}(\varphi) dS$, where $\Gamma(\varphi)$ is the thermal inertia at a given latitude $\varphi$,
and $dS$ is the surface element. The cosine is squared because both the size of the surface element and 
the projection of its normal onto the orbital plane are proportional to the cosine.

Accordingly, a "global" averaged thermal inertia relevant to the Yarkovsky effect ($TI_r$) could be obtained
from the following equation:
\begin{equation}
TI_r \int_{-\frac{\pi}{2}}^{\frac{\pi}{2}}\cos^2 \varphi d\varphi=\int_{-\frac{\pi}{2}}^{\frac{\pi}{2}}TI\left(\varphi\right)\cos^2 \varphi d\varphi
\end{equation}

If we assume linear variation of TI with latitude, then the ratio of $TI_r$ and average thermal inertia $TI_a$ results in
\begin{equation}
    \frac{TI_r}{TI_a}=\frac{\frac{1}{2}\left(TI_e+TI_p\right)+\frac{2}{\pi^2}\left(TI_e-TI_p\right)}{TI_p+\frac{2}{\pi}\left(TI_e-TI_p\right)}
\end{equation}
where $TI_e$ and $TI_p$ are thermal inertia at the equator and pole, respectively. In the case of Bennu, based on the results by \citeauthor{2020SciA....6.3699R}, this translates into a small decrease of our thermal inertia estimate of about 1.2\%, or maximum 2\%, given that spatial variation in thermal inertia at Bennu is not linear from the equator to the pole. Considering this effect, our best estimate of the Bennu surface thermal inertia is 374$_{-53}^{+54}$~J m$^{-2}$ K$^{-1}$ s$^{-1/2}$. The corresponding thermal inertia distribution is shown in Fig.~\ref{fig:bennu_ti_dis}. Despite being shifted towards somewhat larger values, the result is fully compatible with the literature estimates listed in Table~\ref{tab:bennu_TI}, with all terms within 1–$\sigma$ of each other. 

Finally, though \citet{2020SciA....6.3699R} found only small thermal inertia variations with heliocentric distance at Bennu, and we did not consider it in our nominal model, we have investigated how applying our variable thermal inertia model would change the result. Therefore, assuming in the ASTERIA model that thermal inertia scales with heliocentric distance as $r^{-0.75}$, we found a value of $\Gamma$ = 366$_{-52}^{+53}$~J m$^{-2}$ K$^{-1}$ s$^{-1/2}$. This result is even closer to the reference value of Bennu's thermal inertia, which might suggest that variable thermal inertia plays some role for Bennu. Nevertheless, as we do not know how representative the thermal inertia variation exponent of $-0.75$ is in the case of Bennu, we did not consider this result our best match.

\subsection{Model testing - further verification on the Near-Earth asteroids}

To estimate thermal inertia with the ASTERIA model, several input parameters should be provided.
Along with absolute magnitude and orbital parameters, obviously, all the objects must have a Yarkovsky detection.\footnote{We have also assumed that the light curve amplitudes are known and applied the non-sphericity corrections as described in Section~\ref{ss:non-sphericity}. }
The distributions of the other parameters could be, in principle, obtained from the population models. However, as noted by \citet{Fenucci_et_al_AA2023}, if no other parameter is known, the model is not able to provide reliable results, except in special cases, such are super-fast rotators. For this reason, we require here that additionally, the size, albedo, and rotation period of the objects are known. In this case, the density distribution is determined as described in Section~\ref{s:physParam}, but based on the determined albedo values and Gaussian distribution instead of the albedo obtained using the population model. Therefore, the level of the model reliability analyzed here applies to such cases. For objects where the other parameters are available as well, such as bulk density or obliquity, these are neglected for our model validation purposes.

Tables~\ref{tab:other_NEAs_orb} and \ref{tab:other_NEAs_TI} list the selected NEAs for further verification of the ASTERIA model. The objects are selected so that, on one side, they have reasonably well-determined thermal inertia from TPM, and on the other side, to cover as much as possible variety of orbital and physical properties of NEAs. The Tables~\ref{tab:other_NEAs_orb} and \ref{tab:other_NEAs_TI} also provide all the relevant orbital and physical parameters for the model validation set of asteroids.

\begin{table*}[]
    \caption{The orbital properties of 10 NEAs used for validation of the ASTERIA model.
    Source: JPL-Small-Body Database Lookup.}
    \setlength{\tabcolsep}{2pt}
    \centering
    \begin{tabular}{lcccrrrr}
         \hline
         \hline
Asteroid & Orbit type & $a$ [au] & $e$ & $i$ [deg] & $A_2$ [$10^{-15}$ au d$^{-2}$] &  $\text{d}a/\text{d}t$ [$10^{-4}$ au Myr$^{-1}$]  & S/N  \\
\hline
(1620) Geographos & NEA-Apollo & 1.246 & 0.336 & 13.336 & $-$2.79062 &  $-$1.322 & 2.9 \\
(1685) Toro  & NEA-Apollo & 1.367 & 0.436 & 9.383 & $-$3.33046 & $-$1.654  & 5.2 \\
(1862) Apollo & NEA-Apollo & 1.470 & 0.560 & 6.353 & $-$3.78439 & $-$1.948  & 12.0 \\
(1865) Cerberus & NEA-Apollo & 1.080 & 0.467 & 16.102 & $-$10.22195 & $-$4.511  & 2.6 \\
(25143) Itokawa & NEA-Apollo & 1.324 & 0.280 & 1.621  & $-$27.70331 & $-$13.538 & 3.3 \\ 
(29075) 1950 DA & NEA-Apollo & 1.698 & 0.508 & 12.170 & $-$6.87595 & $-$3.805  & 5.1\\
(33342) 1998 WT24 & NEA-Aten & 0.719  & 0.418 & 7.368  & $-$25.25915 & $-$9.094  & 8.5\\
(99942) Apophis & NEA-Aten & 0.923 & 0.191 & 3.339 & $-$29.01086 & $-$11.834 & 149.4\\
(161989) Cacus & NEA-Apollo & 1.123 & 0.214 & 26.065 & $-$10.59405 & $-$4.767 & 5.7 \\
(175706) 1996 FG3 & NEA-Apollo & 1.054 & 0.350 & 1.972 & $-$12.03497 & $-$5.247  & 6.5\\
    \hline
    \end{tabular}
    \label{tab:other_NEAs_orb}    
\end{table*}

\begin{figure}[!ht]
    \centering
    \includegraphics[width=0.5\textwidth, angle=-90]{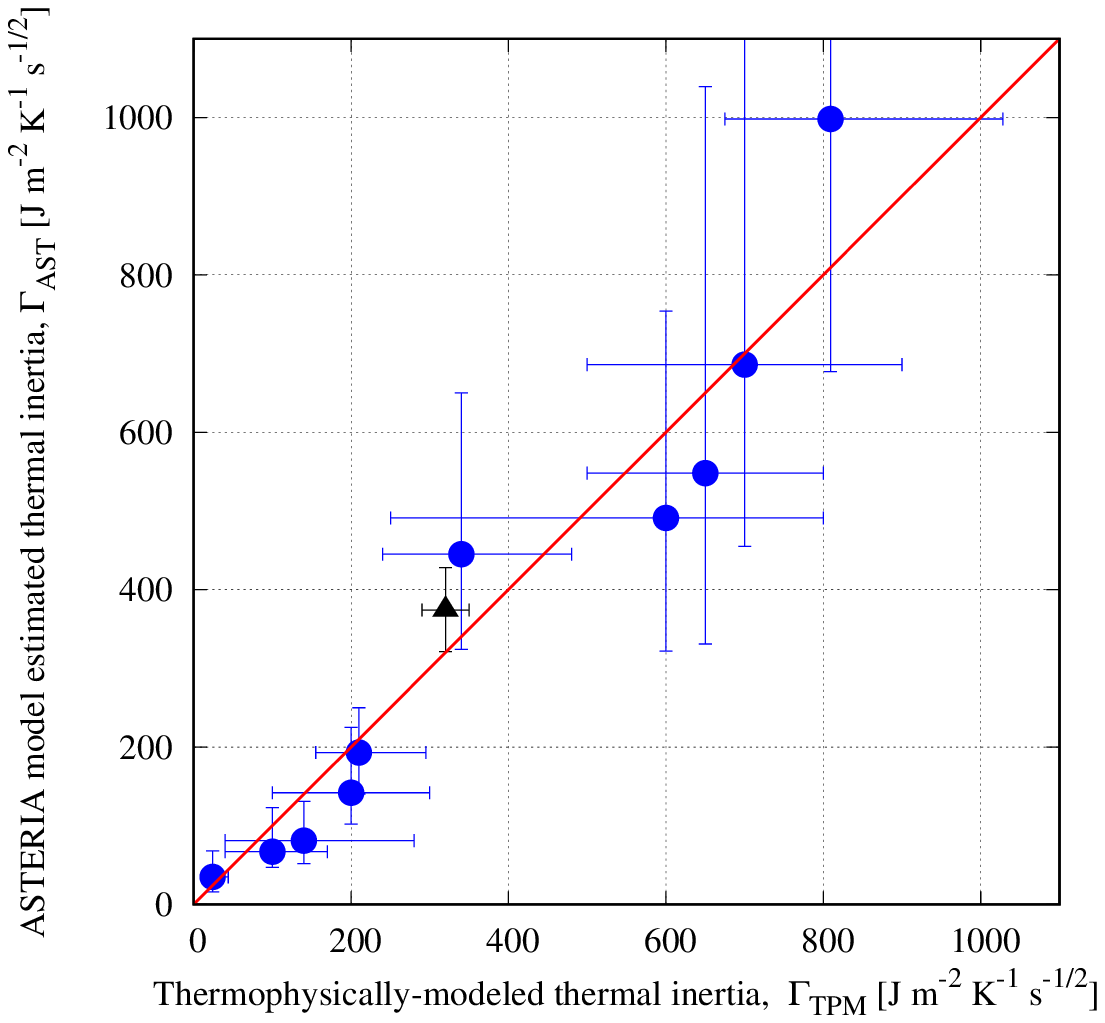}
    \caption{Thermal inertia values derived from thermophysical modeling listed in the literature compared to values derived from the ASTERIA model. The blue points are the results for ten NEAs used for model validation purposes, while the black triangle shows the result for Bennu. The red line is the line of equality where the same results should appear.}
    \label{fig:TI_comparison}
\end{figure}

The ASTERIA asteroid thermal inertia estimator gives values that are generally in good agreement
with those derived from thermophysical modeling (see Fig.~\ref{fig:TI_comparison} and Table~\ref{tab:other_NEAs_TI}). As discussed below for individual cases, all the results obtained with our model are statistically compatible with the corresponding reference literature values. 

Below, we briefly discuss the matching on a case-by-case basis. 

\textbf{(1620) Geographos:} This object has been the subject of thermophysical modelling by \citet{rozitis-green_2014}, and has trustworthy determined thermal inertia. Therefore, we include this object in the validation list. On the other hand, we note that its Yarkovsky detection has S/N of only $\approx$2.9, which is typically considered as marginal detection \citep[e.g.][]{2018A&A...617A..61D}. According to \citet{rozitis-green_2014}, the thermal inertia of Geographos is $340^{+140}_{-100}$ J m$^{-2}$ K$^{-1}$ s$^{-1/2}$. Our model with nominal Yarkovsky drift from the JPL suggests a higher thermal inertia of $\sim$500 J m$^{-2}$ K$^{-1}$ s$^{-1/2}$, with large uncertainties, primarily due to large uncertainty in Yarkovsky drift rate estimation. Nevertheless, the result obtained by \citet{rozitis-green_2014} is still within the 1-$\sigma$ error limits of our estimate, and we consider two results to be roughly consistent.

A bit larger S/N of $\sim 3.6$ of the Yarkovsky detection for Geographos has been obtained by \citet{2022Icar..38315040D}, suggesting slightly faster Yarkovsky drift of $\text{d}a/\text{d}t=-1.34\pm 0.37$ [$10^{-4}$ au Myr$^{-1}$]. Plugging this value into the ASTERIA model, we found $\Gamma = 445^{+205}_{-121}$, which is fully in agreement with estimate from \citeauthor{rozitis-green_2014}.

\textbf{ (1685) Toro: } This is the largest asteroid in our sample, with a diameter of about 3.8~km, but with solid detection of the Yarkovsky effect (S/N$\approx$6.6). An estimated thermal inertia of 210$^{+85}_{-55}$ J m$^{-2}$ K$^{-1}$ s$^{-1/2}$ has been recently given by \citet{2022PSJ.....3...56H}, which is in perfect agreement with our result of 193$^{+57}_{-53}$ J m$^{-2}$ K$^{-1}$ s$^{-1/2}$.

\textbf{ (1862) Apollo:} Its thermal inertia has been estimated to be $140^{+140}_{-100}$ J m$^{-2}$ K$^{-1}$ s$^{-1/2}$ by \citet{2013A&A...555A..20R}. Our nominal result of $81^{+50}_{-29}$ J m$^{-2}$ K$^{-1}$ s$^{-1/2}$ is somewhat lower than the one derived by \citeauthor{2013A&A...555A..20R}. Adopting the physical parameters (geometric albedo $p_v = 0.20\pm0.02$, and diameter $D = 1.55\pm0.07$~km) obtained along with the thermal inertia by \citet{2013A&A...555A..20R}, our model gives slightly higher thermal inertia of $86^{+51}_{-30}$ J m$^{-2}$ K$^{-1}$ s$^{-1/2}$, indicating that the difference in the input parameters is not the main source of the observed difference in the results.
Nevertheless, given the significant uncertainties of both results, we consider the two results consistent. We should underline, however, that for this object, the ASTERIA model also provides the alternative solution of higher thermal inertia in the range of 400$-$650 J m$^{-2}$ K$^{-1}$ s$^{-1/2}$, which we cannot reject based on the available data. Based on the Yarkovsky effect detection algorithm and interdependence between the bulk density and thermal inertia, \citet{farnocchia-etal_2013} suggested that the TI of (1862) Apollo should be in the range 400$-$1000 J m$^{-2}$ K$^{-1}$ s$^{-1/2}$. Therefore, asteroid (1862) Apollo might have higher TI, in line with our alternative solution.

\textbf{ (1865) Cerberus: } The Yarkovsky detection at JPL has S/N of only about 2.6, translating into large uncertainties of our thermal inertia estimate. Nevertheless, our value of 998$^{+625}_{-321}$ J m$^{-2}$ K$^{-1}$ s$^{-1/2}$ agrees with the value of 809$^{+219}_{-134}$ J m$^{-2}$ K$^{-1}$ s$^{-1/2}$, found by \citet{2022PSJ.....3...56H}. We underline that in the case of the low Yarkovsky detection S/N, the results should be taken with caution. For instance, Yarkovsky detection for (1865) Cerberus has also been provided by the \citet{greenberg-etal_2020}. These authors found the semi-major axis drift of $-3.75\pm1.8 \times 10^{-4}$~au~Myr$^{-1}$, therefore with even lower S/N od 2.1. With the Yarkovsky detection from \citeauthor{greenberg-etal_2020}, our model gives the thermal inertia of 1149$^{+961}_{-426}$ J m$^{-2}$ K$^{-1}$ s$^{-1/2}$, formally in agreement with the estimate from \citeauthor{2022PSJ.....3...56H}.

\textbf{ (25143) Itokawa: } This asteroid has one of the largest known thermal inertia values of $700\pm200$ J m$^{-2}$ K$^{-1}$ s$^{-1/2}$ \citep{2014PASJ...66...52M}. Though with large uncertainties caused primarily by a small S/N of 3.3 in the Yarkovsky detection, our nominal result of about 686 J m$^{-2}$ K$^{-1}$ s$^{-1/2}$ perfectly matches the literature value.

\textbf{ (29075) 1950 DA: } \citet{rozitis-etal_2014} found a low thermal inertia of only 24$^{+20}_{-14}$ J m$^{-2}$ K$^{-1}$ s$^{-1/2}$, fully in agreement with our estimate of 35$^{+33}_{-19}$ J m$^{-2}$ K$^{-1}$ s$^{-1/2}$.    

\textbf{ (33342) 1998 WT24: } Based on the thermal-infrared observations and thermophysical modeling \citet{2007Icar..188..414H} found thermal inertia of $200\pm100$ J m$^{-2}$ K$^{-1}$ s$^{-1/2}$, that our nominal result of 142$^{+83}_{-40}$ J m$^{-2}$ K$^{-1}$ s$^{-1/2}$ matches very well.

\textbf{ (99942) Apophis: } For this object, we found four thermal inertia estimates \citep{2014A&A...566A..22M, 2016A&A...585A..10L, 2017RAA....17...70Y, 2022PSJ.....3..124S}, which are not always consistent or are subject to large uncertainties. Nevertheless, we include it in our validation list. Generally speaking, our estimate of $491^{+263}_{-169}$ J m$^{-2}$ K$^{-1}$ s$^{-1/2}$ is consistent with three of the values from the literature, with the only exception being a low value of $100^{+100}_{-52}$ J m$^{-2}$ K$^{-1}$ s$^{-1/2}$ obtained by \citet{2017RAA....17...70Y}.

\textbf{ (161989) Cacus: } Thermal inertia of $548^{+491}_{-169}$ J m$^{-2}$ K$^{-1}$ s$^{-1/2}$ estimated by the ASTERIA model match very well the literature value of $650\pm150$ J m$^{-2}$ K$^{-1}$ s$^{-1/2}$ \citep{2022PSJ.....3...56H}. Despite a good S/N value of 5.7 in the Yarkovsky detection, our result comes with significant uncertainties caused by uncertainties in the size and albedo of the asteroid Cacus. 

\textbf{(175706) 1996 FG3: } There are two available thermal inertia estimates for this object. \citet{2011MNRAS.418.1246W} obtained $120\pm50$ J m$^{-2}$ K$^{-1}$ s$^{-1/2}$, while \citet{2014MNRAS.439.3357Y} found a bit lower value of $80\pm40$ J m$^{-2}$ K$^{-1}$ s$^{-1/2}$. Our nominal result agrees with those estimates, with the mean value being somewhat smaller and in better agreement with the one from \citet{2014MNRAS.439.3357Y}. Using the size and albedo from \citet{2011MNRAS.418.1246W}, our model gives the thermal inertia of
98$^{+59}_{-37}$ J m$^{-2}$ K$^{-1}$ s$^{-1/2}$, which is in full agreement with the value found by those authors.

The above-presented validation results showed that the ASTERIA model is reliable and provides consistent estimates of the thermal inertia with the values available in the literature. The prerequisite is that the diameter and albedo estimates are available for a given object along with the Yarkovsky effect detection. The accuracy of the results obviously depends on the quality of the input parameters, with the model being more sensitive to the Yarkovsky rate and diameter estimations than to the albedo. Nevertheless, as long as the same (or similar values) of the input parameters have been used, the ASTERIA model matches very well the other thermal inertia estimates, and it can be considered a good alternative to thermophysical modeling in proper circumstances.

\begin{table*}[]
    \caption{The physical properties of 10 NEAs used to validate the ASTERIA model. In the last two columns, reference literature values of thermal inertia ($\Gamma_{\textrm{TPM}}$) and our estimates ($\Gamma_{\textrm{AST}}$) are given. In each case, our results refer to the median values and their corresponding 1-$\sigma$ lower and upper spreads.}
    \setlength{\tabcolsep}{2pt}
    \centering
    \begin{tabular}{lcccccccc}
         \hline
         \hline
Asteroid &  $H$    & $D$   & $p_V$ & $\Delta m$  &  $P$  &  $\Gamma_{\textrm{TPM}}$                         & $\Gamma_{\textrm{AST}}$  \\
         &   [mag] &  [km] &   -   & [mag]       & [h]   &  [J m$^{-2}$ K$^{-1}$ s$^{-1/2}$] & [J m$^{-2}$ K$^{-1}$ s$^{-1/2}$]   \\
	\hline
 (1620) Geographos    & 15.32  & 2.46$\pm$0.03    & 0.168$\pm$0.017   & 1.93 & 5.222 & 340$^{+140}_{-100}$ & 445$^{+205}_{-121}$ \\
 (1685) Toro          & 14.31  & 3.810$\pm$0.049 & 0.247$\pm$0.049  & 1.40 & 10.1995 & 210$^{+85}_{-55}$  &  193$^{+57}_{-53}$    \\
 (1862) Apollo        & 16.11  & 1.395$\pm$0.042    & 0.287$\pm$0.041 & 1.15 &  3.066 & 140$^{+140}_{-100}$ &  81$^{+50}_{-29}$  \\
 (1865) Cerberus      & 16.79  & 1.611$\pm$0.013  & 0.136$\pm$0.021   & 2.30  & 6.8039  & 809$^{+219}_{-134}$  &  998$^{+625}_{-321}$    \\
 (25143) Itokawa    	 & 19.26  & 0.33$\pm$0.01     &  0.27$\pm$0.02  & 1.05 & 12.132 & 700$\pm$200 &  686$^{+502}_{-231}$ \\
 (29075) 1950 DA      & 17.28  & 1.810$\pm$0.790      & 0.078$\pm$0.104  & 0.20 &  2.1216 & 24$^{+20}_{-14}$ &  35$^{+33}_{-19}$    \\
 (33342) 1998 WT24	 & 18.02  & 0.432$\pm$0.028  & 0.654$\pm$0.130   & 0.65 &  3.697    & 200$\pm$100  &  142$^{+83}_{-40}$    \\
 (99942) Apophis      & 19.09 & 0.34$\pm$0.04  &  0.35$\pm$0.10  &  1.10 &  30.56 &  600$^{+200}_{-350}$ &   491$^{+263}_{-169}$   \\
(161989) Cacus       & 17.31  & 1.089$\pm$0.244      &  0.229$\pm$0.172     & 1.10 &  3.7538 & 650$\pm$150 &  548$^{+491}_{-217}$   \\
(175706) 1996 FG3    & 18.36  & 1.196$\pm$0.362  & 0.072$\pm$0.039   & 0.20 & 3.5942 & 100$^{+70}_{-60}$  &  67$^{+56}_{-30}$    \\
        \hline
    \end{tabular}
    \label{tab:other_NEAs_TI}

\tablenotetext{~The values of absolute magnitudes $H$ are from the JPL, diameters $D$ and albedos $p_V$ are from \citet{2019PDSS..251.....M}, except for (1620) Geographos \citep{rozitis-green_2014}, (25143) Itokawa \citep{2006Sci...312.1330F,2018Icar..311..175T}, (29075) 1950 DA \citep{2021PSJ.....2..162M}, (99942) Apophis \citep{2018Icar..300..115B}, and (161989) Cacus \citep{2020PSJ.....1....9M}. The values of maximum light curve amplitudes $\Delta m$ and rotation periods $P$ are taken from LCDB \citep{warner-etal_2009}. The reference literature values of thermal inertia $\Gamma_{TPM}$ are, in order of appearance in the table from top to bottom, from [1] \citet{rozitis-green_2014}, [2] \citet{2022PSJ.....3...56H}, [3] \citet{2013A&A...555A..20R}, [4] \citet{2022PSJ.....3...56H}, [5] \citet{2014PASJ...66...52M}, [6] \citet{rozitis-etal_2014}, [7] \citet{2007Icar..188..414H}, [8] \citet{2014A&A...566A..22M}, [9] \citet{2022PSJ.....3...56H}, [10] \citet[][]{2011MNRAS.418.1246W, 2014MNRAS.439.3357Y}. In the case of asteroid (175706) 1996 FG3, an average of the two literature values have been used.}  

\end{table*}

\subsection{ASTERIA model application and criticality of input parameters - an example of asteroid 152671 (1998 HL3)}

After verifying our model in the previous sub-sections using the well-studied asteroid Bennu, and ten additional well-studied NEAs, here we applied the model to the near-Earth asteroid (152671) 1998 HL3 (hereafter HL3) as a case study of a more general situation. 

Contrary to the case of Bennu, we know relatively little regarding the physical parameters of the asteroid HL3.
Its known orbital and physical properties relevant to the model, including those determined in this work, are summarised in Table~\ref{tab:HL3}.

\begin{table*}[]
    \caption{Orbital and physical parameters of asteroid (152671) 1998 HL3.}
    \centering
    \begin{tabular}{lcc}
         \hline
         \hline
    Parameter & Value & Reference \\
         \hline
    Semi-major axis, $a$   & 1.128978 au   & NEOCC - Epoch 60200.0 MJD  \\
    Eccentricity, $e$  &  0.366000  &  NEOCC - Epoch 60200.0 MJD \\
    Inclination, $i$  &  2.6800 deg  &  NEOCC - Epoch 60200.0 MJD \\
    $A_2$ acceleration, $A_2$  & $-49.68592$ [$10^{-15}$ au d$^{-2}$]  &  NEOCC - Epoch 60200.0 MJD \\
    Semi-major axis drift, $da/dt$  & $-22.4189\pm3.0116$ $10^{-4}$ au Myr$^{-1}$ & NEOCC - Epoch 60200.0 MJD \\
    Yarkovsky detection S/N  & 7.4  &  NEOCC - Epoch 60200.0 MJD \\
    \hline
    Absolute magnitude, $H$   & 20.184 & NEOCC \\
    Diameter, $D$  & 298 $\pm$ 7  m & \citet{2011ApJ...741...68M} \\
    Albedo, $p_V$   & 0.200 $\pm$ 0.037 & \citet{2011ApJ...741...68M} \\
    Max light curve amplitude, $\Delta m$ & 0.15  & this work \\
    Rotation period, $P$  & 5.3 h  & this work (preliminary) \\
    \hline
    Thermal inertia, $\Gamma$   &  506$^{+228}_{-121}$ J m$^{-2}$ K$^{-1}$ s$^{-1/2}$ & this work (see text for alternative solutions) \\
       \hline
    \end{tabular}
    \label{tab:HL3}
\end{table*}

We also used this subsection to highlight how important is the determination of rotation periods for our model. In this respect, we present our photometric observation of asteroid HL3, aiming to obtain its light curve and derive its amplitude and rotation period. 

\subsubsection{Observations from Astronomical Station Vidojevica}

\begin{table*}[]
    \caption{Observational circumstances of 52671 (1998 HL3).}
    \centering
    \begin{tabular}{lcccccccc}
         \hline
         \hline
UT Date & Exp. time [s] & No. of exposures & Filter & Avg. seeing (FWHM) [''] & $r_h $ [au] & $\Delta$ [au] & $\alpha $ [deg]  & $m(r_h, \Delta, \alpha )$ \\
	\hline
	2022 May 4  & 200 & 76 (28) & R & 1.2 & 1.268  & 0.265 & 9.80 & 18.37 \\
	2022 May 19 & 200 & 55 (44) & R & 1.7 & 1.189 & 0.216 & 31.56 & 18.49 \\
	2022 May 20 & 200 & 67 (51) & R & 1.6 & 1.184 & 0.214 & 33.39 & 18.51 \\
        \hline
    \end{tabular}
    \label{tab:HL3_obs}

\tablenotetext{~In the "No. of exposure" column, the numbers in brackets show how many images from each night were used for the period determination. } 

\end{table*}

Observations were collected on the 5th of May 2022, as well as on the 19th and 20th of May 2022 from the Astronomical station Vidojevica (MPC code C89), using the 1.4 m Milankovi\'c telescope. We used an Andor iKon-L 2024$\times$2024 pixel CCD camera, which has a field of view of 13.3$\times$13.3 arcmin and a pixel size of 13.5$\times$13.5 $\mu$m. All observations were made in the Standard Johnson-Cousin R-filter. Light curve construction and rotational period determinations were done in MPO Canopus software version 10.8.6.3 \citep{Warner2021}. Period determination was performed using the implemented Fourier Analysis of Light Curves (FALC) algorithm. 

\begin{figure}[!ht]
    \centering
    \includegraphics[width=0.48\textwidth]{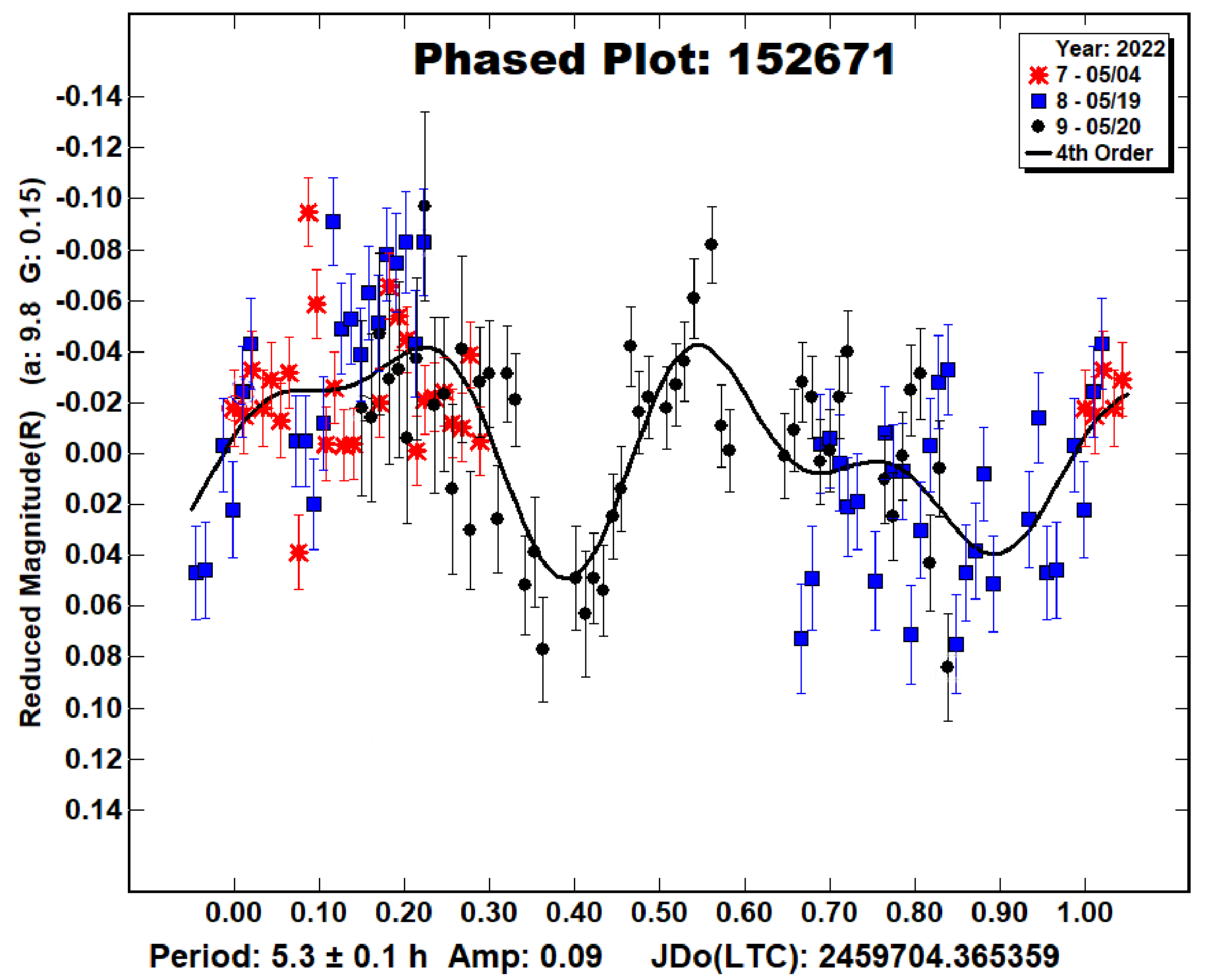}
    \caption{Data for two nights photometric observations of asteroid (152671) 1998 HL3 and the corresponding light curve for the nominal period solution of $P=5.3$~h, plotted with MPO Canopus.}
    \label{fig:152671_rot_per}
\end{figure}

The obtained light curve of HL3 is shown in Figure~\ref{fig:152671_rot_per}. As the object was relatively faint (visual magnitude of 18.4-18.5), the data is noisy, and a unique period solution cannot be reliably derived.
The nominal period solution is 5.3~h, but the alternative solutions of 2.6 and 1~h can not be ruled out. Therefore, additional observations of HL3 are needed to establish its rotation period firmly. 

The amplitude of our nominal light curve solution is $\Delta m=0.1$ mag. Given that $\Delta m$ depends on the phase angle \citep{1990A&A...231..548Z,2006A&A...454..367G}, a somewhat larger amplitude is possible as well, but likely not much larger as we observed at phase angle $\alpha$ of $\sim$32 degrees. To account for the object's non-sphericity in our model, we adopted an amplitude of $\Delta m$=0.15. Following the methodology described in Section~\ref{ss:non-sphericity}, for the asteroid HL3, we found an axes ratio of $a/c = 1.15$. These yield a non-sphericity correction factor of $\xi=0.91$.

\subsubsection{Thermal inertia estimation of asteroid (152671) 1998 HL3}

Though the obtained period solution is still unreliable and needs further verification, we obtained a preliminary estimate of the thermal inertia of HL3. We also take the opportunity to highlight once again the importance of an accurate period solution for reliable thermal inertia estimations with the ASTERIA model.
Additionally, we demonstrated how the results may change when the variable thermal inertia model is used.

\begin{figure}[!ht]
    \centering
    \includegraphics[width=0.48\textwidth]{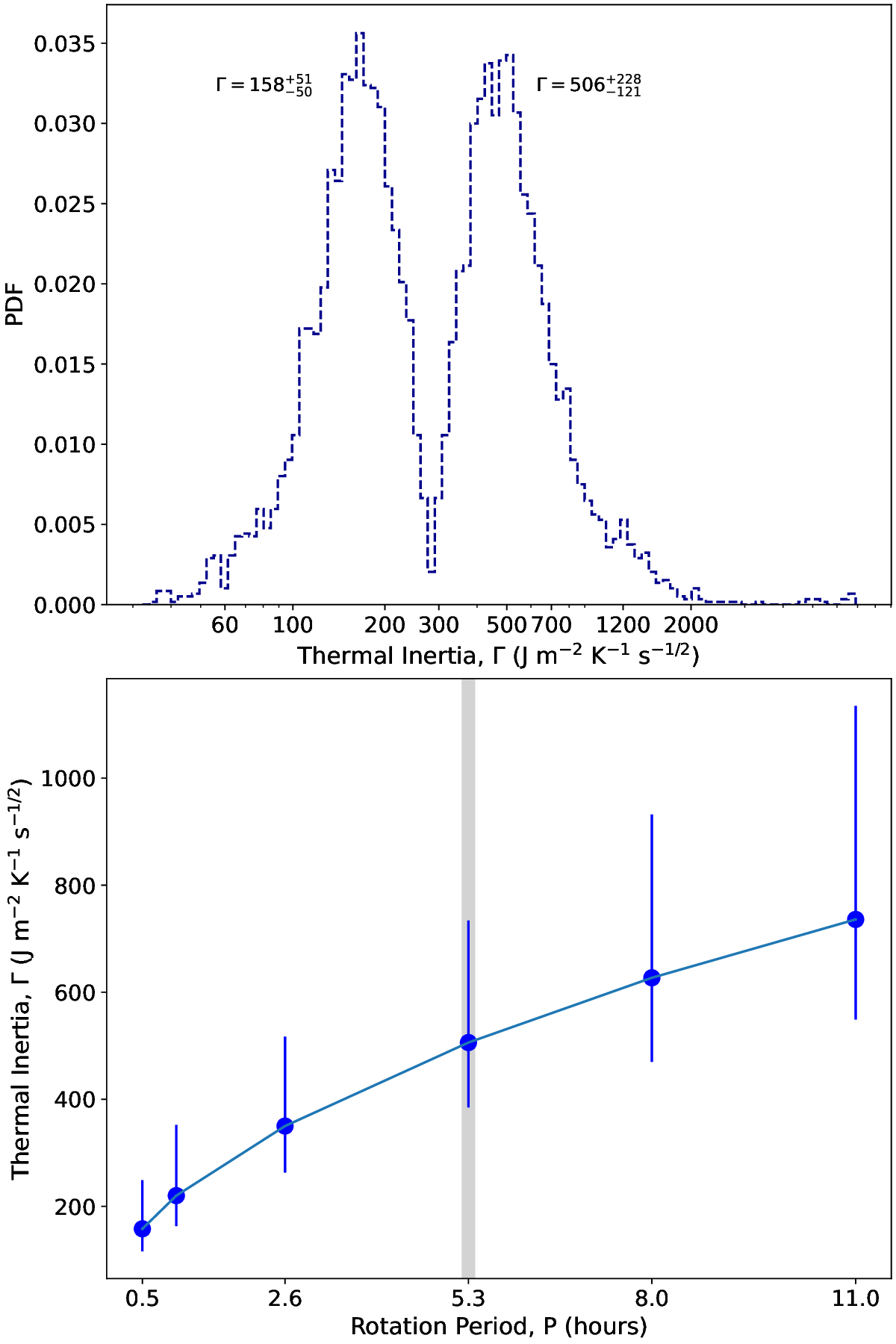}
    \caption{The thermal inertia estimates for asteroid (152671) 1998 HL3 generated with the ASTERIA model.
    The upper panel shows the full distribution obtained for our nominal rotation period solution. The lower panel shows how the results change with the rotation period, highlighting the importance of reliable determination of this parameter. The median values of $\Gamma$ and their corresponding lower and upper $1-\sigma$ uncertainties are shown. The vertical grey area in the lower panel indicates the result for the nominal period solution. See text for additional details.}
    \label{fig:HL3_ti_dis}
\end{figure}

In the case of the HL3, rejecting one of the peaks is not trivial, and none of them could be rejected based on the currently available data. However, given that our aim here is primarily to demonstrate different aspects of the model and the relevance of the input parameters, we adopted the higher peak in our analysis. Adopting the lower peak does not change our overall conclusions.

In the upper panel of Fig.~\ref{fig:HL3_ti_dis}, we show the full thermal inertia distribution generated by our model using the parameters listed in Table~\ref{tab:HL3}. This distribution corresponds to thermal inertia of 506$^{+228}_{-121}$ J m$^{-2}$ K$^{-1}$ s$^{-1/2}$. If, however, we also consider the variable thermal inertia in our model, the result drops to 419$^{+177}_{-98}$ J m$^{-2}$ K$^{-1}$ s$^{-1/2}$. This change is relatively small, though the effect is larger than in the case of Bennu. It implies that the variable thermal inertia plays a limited role for asteroids with orbital eccentricities $e\leq0.35$, with HL3 being a marginal case.\footnote{Throughout this paper, we opt not to generally use the variable thermal inertia model because its results depend on the exponent assumed, which may vary from asteroid to asteroid \citep{rozitis-etal_2018,2021PSJ.....2..161M}. }

The lower panel of Fig.~\ref{fig:HL3_ti_dis} demonstrates how our estimates of the asteroid HL3's thermal inertia vary with the rotation period. The results could change up to a factor of a few with this parameter. Keeping all the other parameters fixed, the TI increases with P, as expected. Therefore, the reliability of thermal inertia results obtained with ASTERIA also depends on the accuracy of the rotation period. This fact was also noted by \citet{2021A&A...647A..61F,Fenucci_et_al_AA2023}, who exploited the fast rotation of two small NEAs 2011~PT and 2016~GE1 to constrain their thermal inertia to low values.

\subsubsection{List of priority asteroids for rotation period determination}

Having demonstrated how the (un)known rotation period affects the thermal inertia estimation by the ASTERIA model
in Table~\ref{tab:observable_objects}, we provide a list of priority targets suitable for observations in the coming years. These objects have all the critical parameters used by the ASTERIA model determined, but the light curve.
Therefore, estimating the amplitude of the light curves and determining the rotation periods would allow evaluation of the thermal inertia of those asteroids.

\begin{table*}[!ht]
    \caption{Priority targets for light curve determination that would allow thermal inertia estimation by the ASTERIA. Objects suitable for observation (visual magnitude $<$ 22, solar elongation $>$ 90$^\circ$) until 2027 are listed. For each object, the month of its brightest visual magnitude has been given.}
    \setlength{\tabcolsep}{2pt}
    \centering
    \begin{tabular}{lccc}
         \hline
         \hline
Asteroid &  Absolute    & Peak visual & Best period  \\
      {}   &   magnitude  &  magnitude & for observation \\
	\hline
 (506590) 2005 XB1    & 21.96  & 21.0 & Nov 2023\\
 (152671) 1998 HL3 & 20.24  & 20.7  & Jan 2024\\
 (164207) 2004 GU9    & 21.19  & 20.7 & Jan 2024\\
 (234341) 2001 FZ57    & 18.95  & 17.3 & Apr 2024\\
 (488803) 2005 GB120    & 20.50  & 19.7 & Apr 2024 \\
 (215588) 2003 HF2    & 19.42  & 20.5 & May 2024\\
 (247517) 2002 QY6    & 19.72  & 20.0 & Sep 2024\\
 (613995) 2008 NP3    & 22.83  & 21.3 & April 2025\\
 (66400) 1999 LT7    & 19.29  & 18.2 & May 2025\\
 (468468) 2004 KH17    & 21.98  & 20.5 & May 2025\\
 (154590) 2003 MA3    & 21.71  & 21.3 & May 2025\\
 (307070) 2002 AV31    & 20.64  & 19.7 & Aug 2025\\
 (162080) 1998 DG16    & 19.95  & 17.5 & Mar 2026\\
 ~~~~~-~~~~2002 JR100    & 24.30  & 20.7 & May 2026\\
 ~~~~~-~~~~2010 FX9    & 24.06  & 19.0 & Sep 2026\\
 (138947) 2001 BA40    & 18.82  & 18.3 & Jan 2027\\
 
        \hline
    \end{tabular}
    \label{tab:observable_objects}    
\end{table*}

\section{New thermal inertia estimations for 38 near-Earth asteroids}
\label{s:new_TI}

As the final step in our analysis, we have selected asteroids from the JPL service for which all the relevant input parameters for the ASTERIA model, including the size and albedo, are available. We found 38 such objects, with all of them being near-Earth asteroids, and estimated their thermal inertia by means of our model. Among those 38 objects, 29 are characterized as potentially hazardous asteroids (PHA), making our results relevant also from the planetary defense point of view. 

\begin{figure}[!ht]
    \centering
    \includegraphics[width=0.5\textwidth, angle=-90]{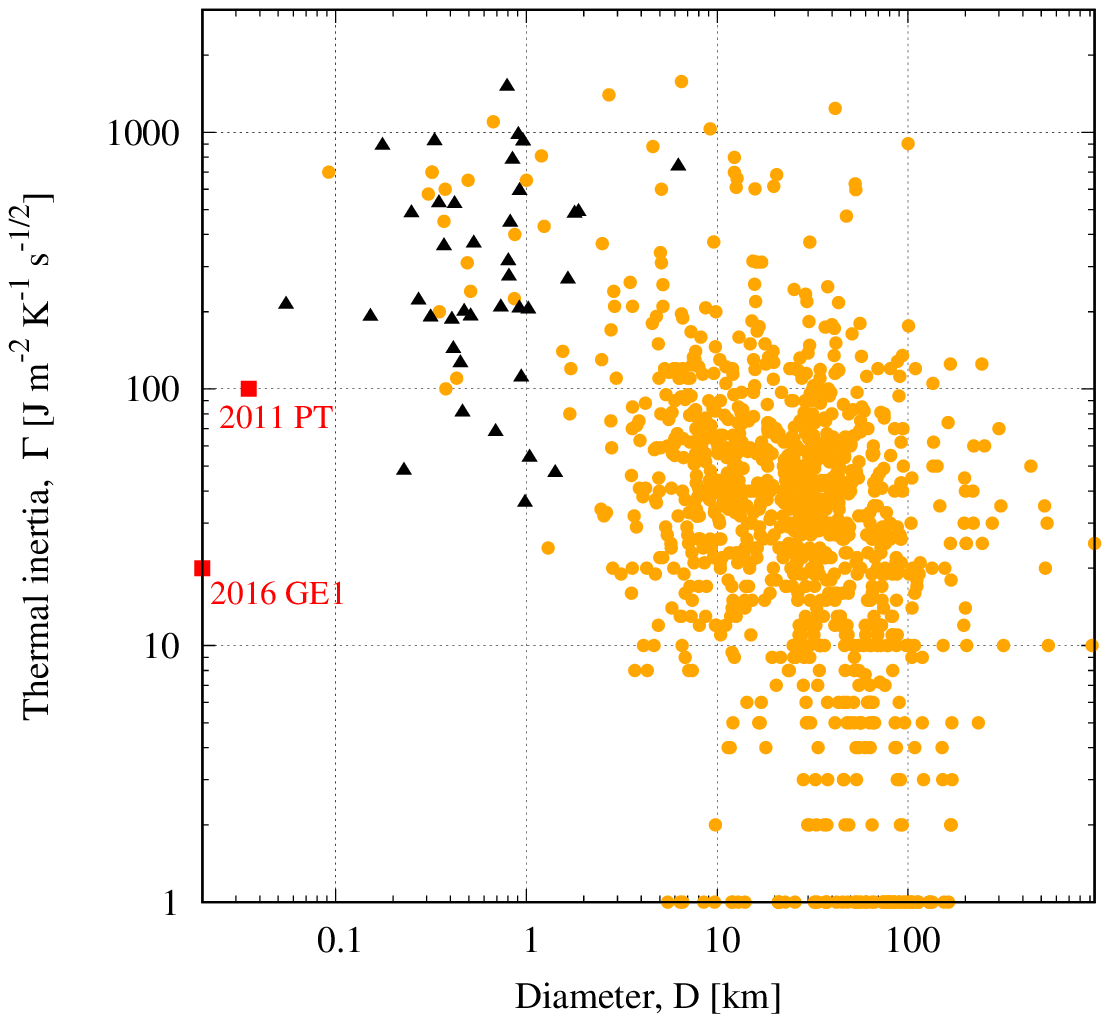}
    \caption{Thermal inertia vs. diameter values for objects with available thermal inertia estimates. Our sample of thermal inertia estimates for 38 asteroids (black triangles) is shown along with the thermophysically derived literature values (orange circles) \citep[see][and references therein]{2021PSJ.....2..161M, 2022PSJ.....3...56H}. Note that numbers from the literature include multiple instances of the same asteroid. Additionally, our two previous estimates for asteroids 2011~PT \citep{2021A&A...647A..61F} and 2016~GE1 \citep{Fenucci_et_al_AA2023} are shown as red squares and indicated in the plot.}
    \label{fig:TI_overview}
\end{figure}

The complete list of objects, relevant input parameters, and thermal inertia estimations are given in Tables~\ref{tab:new_NEAs_orb} and \ref{tab:new_TI}. In Fig.~\ref{fig:TI_overview}, we have shown thermal inertia vs. diameter values for our results and the other available literature estimates. It highlights the ASTERIA model's contribution to thermal inertia estimates of sub-km asteroids. Comments related to some individual cases are given below.

(441987) 2010~NY65 is a very interesting case. Along with the $A_2$ acceleration component, JPL
also found an out-of-orbital plane, the $A_3$ acceleration component. \citet{2023PSJ.....4...35S} hypothesized that the $A_3$ component in small asteroids could be due to non-detected cometary-like activity. Therefore, the same reasoning places the object in the category of "dark comets". Its albedo is 0.071$\pm$0.014 \citep{2011ApJ...741...90M}, in agreement with the albedo of cometary nuclei. The very low TI that we found may further support the link to comets, as values of TI below 100 J m$^{-2}$ K$^{-1}$ s$^{-1/2}$ are common for comets \citep[][]{2019SSRv..215...29G}. However, based on the collected spectrophotometry data, the object was found to be $S$ (or $S_V$) spectral type \citep{2018A&A...615A.127I, 2018P&SS..157...82P}, making its real nature quite puzzling.

(3200) Phaeton is an active asteroid \citep{2012AJ....143...66J}, and the measured drift in the semimajor axis might not be entirely given by the Yarkovsky effect, because the activity could also contribute to the drift. Therefore, our results are based on the assumption that the possible contribution of the Phaeton's activity to its $A_2$ acceleration is negligible. This should be a reasonable assumption, given that the activity is expected to cause radial ($A_1$) or out-of-plane ($A_3$) acceleration components, which are yet to be detected according to JPL data. This object also has available thermal inertia estimates in the literature. The thermal inertia of Phaethon has been estimated to be $\Gamma = 600\pm200$ by \citet{2016A&A...592A..34H} and $\Gamma = 880_{-330}^{+580}$ J m$^{-2}$ K$^{-1}$ s$^{-1/2}$ by \citep{2019AJ....158...97M}. Both results are consistent with our result within $1-\sigma$ uncertainties, suggesting that our model works well even in the case of active asteroids.\footnote{\citet{2022Icar..38815226M} found that Phaethon's thermal inertia increases with temperature across all sightings, possibly indicating the existence of two distinct regions of different surface thermal inertia. As we show in the case of Bennu, the spatial variation across the surface can affect the results obtained with the ASTERIA model. Still, the effect is generally significantly smaller than the uncertainties caused by the input parameters.}

Another quite intriguing case is (1566)	Icarus. With an orbital eccentricity of $e = 0.827$, it is the second most eccentric orbit in our sample. Our estimation suggests a very low TI of asteroid Icarus. Indeed, an alternative solution from the model suggests the TI $>$ 3000, which, however, seems unlikely. It is classified as a Q-type object \citep{binzel-etal_2019}, suggesting the presence of a fresh surface material. The low thermal inertia, therefore, could be due to loosely packed, highly porous, material at the surface. The unweathered material could be exposed at the surface by different processes, such as planetary tidal forces, micrometeorite impacts, or thermal cracking. Though, in principle, any of these mechanisms could be involved, we believe the high temperature during the perihelion passage plays a key role, as perihelion (1566) Icarus is only 0.19~au from the Sun. Primitive C-type near-Earth objects are typically destroyed close to the Sun \citep{granvik-etal_2016}, while rocky asteroids, as Icarus is supposed to be, are more likely to survive. Still, thermal cracking of the surface material could be a frequent process.

(163899) 2003~SD220 is a complex case. We found a very high TI, but several factors may affect this estimate. The asteroid has a very long rotation period of about 285 hours, and therefore, a high thermal inertia may be required for the Yarkovsky effect to work. However, the slow rotation may also suggest that it could be in a tumbling rotation state \citep{2016MPBu...43..143W}. The maximum amplitude of its light curve is also very high, about 2.2~mag, indicating a very elongated shape. In such cases, our non-sphericity scaling approach described in Sec.~\ref{ss:non-sphericity} may not be appropriate. However, we have also obtained the thermal inertia neglecting non-sphericity correction and obtained $974_{-424}^{+713}$ J m$^{-2}$ K$^{-1}$ s$^{-1/2}$. Despite being somewhat lower, this supports the high thermal inertia solution for asteroid (163899) 2003~SD220.

Contrary to the previous example, asteroid 2010~VK139 is a rapid rotator, with a period of only about 108 seconds. Our estimate of its TI of $213_{-107}^{+225}$ J m$^{-2}$ K$^{-1}$ s$^{-1/2}$, potentially makes it another member of a group of small super fast rotators characterized by the relatively low thermal inertia, below $\sim 250$ J m$^{-2}$ K$^{-1}$ s$^{-1/2}$. Still, the TI of VK139 is significantly higher than in the cases of 2011~PT, and especially 2016~GE1 \citep{2021A&A...647A..61F,Fenucci_et_al_AA2023}.

In a more general context, an interesting aspect of thermal inertia investigation is whether typical thermal inertia values could be linked to specific asteroid spectral types (composition). Detailed analysis along these lines is beyond the purpose of this work. We only briefly discuss the case of V-type objects.

Four objects in our sample are classified as V-type asteroids. These are (3908) Nyx, (5604) 1992~FE, (297418) 2000~SP43, and (363599) 2004~FG11 \citep{2014Icar..228..217T,2019Icar..324...41B}. \citet{2020AJ....159..264J} recently studied the thermal inertia of 10 multi-km Vesta family members, presumed to be V-type asteroids, and found the average thermal inertia of 42 J m$^{-2}$ K$^{-1}$ s$^{-1/2}$, with all analyzed 10 objects having $\Gamma \lesssim 100$ J m$^{-2}$ K$^{-1}$ s$^{-1/2}$. Among four V-types studied in our work, three of them have nominal terminal inertia $\Gamma \lesssim 200$ J m$^{-2}$ K$^{-1}$ s$^{-1/2}$ which are consistent with low values found by \citeauthor{2020AJ....159..264J}, taking into account that our estimates refer to near-Earth objects orbiting at smaller heliocentric distance. The asteroid (5604) 1992~FE has also been classified as V-type, but its thermal inertia has been found by \citet{2021PSJ.....2..161M} to be quite high of $1000_{-600}^{+2200}$, in agreement with our nominal result of $922_{-332}^{+769}$ J m$^{-2}$ K$^{-1}$ s$^{-1/2}$. That suggests a large range of possible thermal inertia values also associated with the V-type asteroids, as found in the case for C- and S-type classified objects \citep{2020ApJ...901..140H}. Still, we note that, in our case, an alternative solution for the (5604) 1992~FE thermal inertia of $153_{-71}^{+85}$ J m$^{-2}$ K$^{-1}$ s$^{-1/2}$ cannot be ruled out. If the latter estimate is correct, that would indicate that low thermal inertia values preferentially characterize V-type objects.

\begin{table*}[]
    \caption{The orbital properties of 38 NEAs used in the ASTERIA model for thermal inertia estimations.
    Source: JPL-Small-Body Database Lookup.}
    \setlength{\tabcolsep}{2pt}
    \centering
    \begin{tabular}{llcccrrrr}
         \hline
         \hline
Asteroid & Orbit type & PHA & $a$ [au] & $e$ & $i$ [deg] & $A_2$ [$10^{-15}$ au $d^{-2}$] &  $\text{d}a/\text{d}t$ [$10^{-4}$ au Myr$^{-1}$]  & S/N  \\
\hline
(1566) Icarus      & NEA-Apollo & Yes & 1.078 & 0.827 & 22.799 &  -3.02938 & -1.335  & 4.5 \\
(2062) Aten        & NEA-Aten   & No  & 0.967 & 0.183 & 18.934 & -10.60849 & -4.430  & 9.2 \\
(2063) Bacchus     & NEA-Apollo & No  & 1.078 & 0.349 &  9.432 & -10.30184 & -4.543  & 3.3 \\
(3200) Phaethon    & NEA-Apollo & Yes & 1.271 & 0.890 & 22.273 &  -6.68933 & -3.203  & 8.4 \\
(3361) Orpheus     & NEA-Apollo & Yes & 1.210 & 0.323 &  2.662 & -18.91437 & -8.837  & 7.9 \\
(3908) Nyx         & NEA-Amor   & No  & 1.927 & 0.459 &  2.186 &  12.13284 &  7.153  & 5.4 \\
(4034) Vishnu      & NEA-Apollo & Yes & 1.060 & 0.444 & 11.168 & -40.25166 &-17.595  & 6.0 \\
(4179) Toutatis    & NEA-Apollo & Yes & 2.544 & 0.625 &  0.448 &  -7.71667 & -5.227  & 7.9 \\
(4660) Nereus      & NEA-Apollo & Yes & 1.485 & 0.359 &  1.454 &  13.61057 &  7.043  & 2.1 \\
(5604) 1992 FE     & NEA-Aten   & Yes & 0.928 & 0.406 &  4.715 & -26.53365 &-10.857  & 3.3 \\
(6489) Golevka     & NEA-Apollo & Yes & 2.485 & 0.612 &  2.266 & -12.18508 & -8.157  & 8.4 \\
(7341) 1991 VK     & NEA-Apollo & Yes & 1.841 & 0.506 &  5.409 &  -5.20601 & -3.000  & 2.9 \\
(35107)	1991VH     & NEA-Apollo & Yes & 1.137 & 0.144 & 13.911 &   1.88307 &  0.852  & 0.3 \\
(65679)	1989 UQ    & NEA-Aten   & Yes & 0.915 & 0.265 &  1.303 & -41.30455 &-16.776  & 4.0 \\
(68950) 2002 QF15  & NEA-Apollo & Yes & 1.057 & 0.344 & 25.151 &  -6.39820 & -2.793  & 4.4 \\
(85774)	1998 UT18  & NEA-Apollo & Yes & 1.403 & 0.329 & 13.587 &  -6.85646 & -3.448  & 5.0 \\
(85953)	1999 FK21  & NEA-Aten   & No  & 0.739 & 0.703 & 12.610 & -13.28342 & -4.848  & 9.4 \\
(85990) 1999 JV6   & NEA-Apollo & Yes & 1.008 & 0.311 &  5.359 & -35.66767 &-15.210  &13.5 \\
(152563) 1992 BF   & NEA-Aten   & No  & 0.908 & 0.271 &  7.257 & -25.89494 &-10.476  &27.6 \\
(152754) 1999 GS6  & NEA-Apollo & Yes & 1.191 & 0.497 &  2.020 & -27.93183 &-12.946  & 4.3 \\
(153201) 2000 WO107& NEA-Aten   & Yes & 0.911 & 0.781 &  7.769 & -16.32431 & -6.617  & 4.0 \\
(163348) 2002 NN4  & NEA-Aten   & Yes & 0.876 & 0.434 &  5.420 &  33.34498 & 13.252  & 3.0 \\
(163899) 2003 SD220& NEA-Aten   & Yes & 0.828 & 0.210 &  8.538 & -14.21536 & -5.492  & 3.4 \\
(185851) 2000 DP107& NEA-Apollo & Yes & 1.365 & 0.376 &  8.672 &   9.61206 &  4.769  &15.2 \\
(242191) 2003 NZ6  & NEA-Aten   & No  & 0.793 & 0.492 & 18.245 &  27.77914 & 10.507  & 6.1 \\
(297418) 2000 SP43 & NEA-Aten   & Yes & 0.811 & 0.467 & 10.344 & -21.62628 & -8.273  & 8.1 \\
(337248) 2000 RH60 & NEA-Aten   & No  & 0.826 & 0.551 & 19.651 & -35.36084 &-22.830  & 5.1 \\
(363027) 1998 ST27 & NEA-Aten   & Yes & 0.819 & 0.530 & 21.064 &  15.69986 &  6.035  &11.4 \\
(363505) 2003 UC20 & NEA-Aten   & Yes & 0.781 & 0.337 &  3.811 &  -6.55193 & -4.473  & 4.9 \\
(363599) 2004 FG11 & NEA-Apollo & Yes & 1.587 & 0.724 &  3.131 & -55.14249 &-29.497  & 9.1 \\
(385186) 1994 AW1  & NEA-Amor   & Yes & 1.105 & 0.075 & 24.098 &  14.25635 &  6.364  & 7.4 \\
(398188) Agni      & NEA-Aten   & Yes & 0.864 & 0.273 & 13.252 &  -3.20624 & -1.266  & 3.2 \\
(410777) 2009 FD   & NEA-Apollo & No  & 1.164 & 0.493 &  3.126 &  83.82127 & 38.402  & 5.7 \\
(422686) 2000 AC6  & NEA-Aten   & Yes & 0.854 & 0.286 &  4.699 &  41.61499 & 16.328  & 6.8 \\
(441987) 2010 NY65 & NEA-Apollo & Yes & 1.003 & 0.370 & 11.546 & -38.35183 &-16.312  &26.0 \\
(443880) 2001 UZ16 & NEA-Apollo & Yes & 1.760 & 0.427 & 12.673 & -52.37844 &-29.508  & 5.1 \\
(511684) 2015 BN509& NEA-Apollo & Yes & 1.007 & 0.568 &  4.153 &  82.09613 & 34.979  & 3.7 \\
~~-~~~~2010 VK139  & NEA-Aten   & No  & 0.780 & 0.282 & 26.959 & -98.15438 &-36.810  & 2.6 \\
    \hline
    \end{tabular}
    \label{tab:new_NEAs_orb}    
\end{table*}

\begin{table*}[]
    \caption{New thermal inertia estimates from the ASTERIA model for 38 NEAs, along with the physical input parameters used. The reported values of the TI are median values and their associated $1-\sigma$ lower and upper uncertainties. For all asteroids, we reported nominal values ($\Gamma_{\textrm{nominal}}$) of thermal inertia. In some cases where it was not possible to reliably discriminate between the two solutions, we have also reported our alternative solution ($\Gamma_{\textrm{alternative}}$) (the second peak in the distribution of TI).}
    \setlength{\tabcolsep}{2pt}
    \centering
    \begin{tabular}{lccccccccc}
         \hline
         \hline
Asteroid &  $H$    & $D$   & $p_V$ & $\Delta m$  &  $P$  &  $\Gamma_{\textrm{nominal}}$    & $\Gamma_{\textrm{alternative}}$ & Notes  \\
         &   [mag] &  [km] &   -   & [mag]       & [h]   &  [J m$^{-2}$ K$^{-1}$ s$^{-1/2}$] & [J m$^{-2}$ K$^{-1}$ s$^{-1/2}$] & \\
	\hline
(1566)	Icarus    & 16.50  & 1.417$\pm$0.123  & 0.199$\pm$0.110  & 0.20  &  2.273 & $47_{-21}^{+53}$  & -  &   \\
(2062)	Aten      & 17.11  & 0.804$\pm$0.031  & 0.520$\pm$0.101  & 0.25  &  42.15 & $315_{-105}^{+160}$  & -  &   \\
(2063) Bacchus    & 17.25  & 1.024$\pm$0.020 & 0.203$\pm$0.022 & 0.42 & 14.904 & $204_{-83}^{+115}$  & $1425_{-556}^{+1444}$ &    \\
(3200)  Phaethon  & 14.40  & 6.25$\pm$0.15    & 0.107$\pm$0.011   & 0.30  & 3.604 & $737_{-347}^{+784}$  & - &    \\
(3361)	Orpheus   & 19.39  & 0.348$\pm$0.055  & 0.357$\pm$0.135  & 0.30 & 3.533  & $529_{-199}^{+312}$  & $60_{-22}^{+36}$  &    \\
(3908) Nyx        & 17.46  & 1.04$\pm$0.16  & 0.16$\pm$0.08 & 0.46  & 4.426 & $54_{-20}^{+30}$  & $352_{-125}^{+204}$ &  \\
(4034) Vishnu     & 18.52  & 0.42$\pm$0.05  & 0.52$\pm$0.15   & 0.62  & 44.4 & $525_{-161}^{+199}$  & $1970_{-568}^{+1052}$  &  \\
(4179)	Toutatis  & 15.32  & 1.788$\pm$0.376  & 0.405$\pm$0.137   & 1.20  & 176.0 & $482_{-165}^{+227}$  & - &    \\
(4660)	Nereus    & 18.65  & 0.33$\pm$0.05 & 0.55$\pm$0.17 & 1.00  & 15.175 & $924_{-403}^{+814}$  & $69_{-36}^{+58}$  &  \\
(5604)	1992 FE   & 17.38  & 0.962$\pm$0.011 & 0.527$\pm$0.096 & 0.20 & 5.337 & $922_{-332}^{+769}$  & $153_{-71}^{+85}$  &  \\
(6489)	Golevka   & 19.22  & 0.53$\pm$0.03  & 0.151$\pm$0.023   & 1.00 & 6.026 & $369_{-105}^{+204}$  & $93_{-34}^{+38}$ &   \\
(7341)	1991 VK   & 16.96  & 0.982$\pm$0.316  & 0.235$\pm$0.107   & 0.70  &  4.210    & $36_{-20}^{+38}$  & $852_{-461}^{+2073}$    \\
(35107)	1991VH    & 16.91  & 0.908$\pm$0.035  & 0.408$\pm$0.048 & 0.20 & 2.624 & $978_{-583}^{+2096}$  & $21_{-14}^{+30}$ &   \\
(65679)	1989 UQ   & 19.62  & 0.918$\pm$0.010 & 0.033$\pm$0.007 & 0.36 & 7.746 & $206_{-73}^{+91}$  & - &    \\
(68950) 2002 QF15 & 16.39  & 1.650$\pm$0.555  & 0.178$\pm$0.077 & 0.36  & 45.24 & $267_{-137}^{+233}$  & -  &  \\
(85774)	1998 UT18 & 19.17  & 0.939$\pm$0.007  & 0.042$\pm$0.007  & 0.86  & 55. & $111_{-40}^{+95}$  & - &   \\
(85953)	1999 FK21 & 18.15  & 0.922$\pm$0.207  & 0.122$\pm$0.058  & 0.60  &  28.08 & $591_{-341}^{+838}$  & -  &   \\
(85990) 1999 JV6  & 20.27  & 0.451$\pm$0.026  & 0.095$\pm$0.023  & 0.90  &  6.538 & $126_{-42}^{+73}$  & $1145_{-431}^{+635}$   & \\
(152563) 1992 BF  & 19.81  & 0.272$\pm$0.077  & 0.287$\pm$0.189  & 0.60  &  32.0  & $221_{-100}^{+185}$  & -  &   \\
(152754) 1999 GS6 & 19.30  & 0.414$\pm$0.077 & 0.216$\pm$0.105 & 0.20 &8.021 & $143_{-67}^{+104}$  &  $1481_{-657}^{+1906}$ &  \\
(153201) 2000 WO107& 19.34 & 0.510$\pm$0.083 & 0.129$\pm$0.058 & 1.24 & 5.022 & $192_{-107}^{+257}$  & - &    \\
(163348) 2002 NN4  & 20.08 & 0.735$\pm$0.243 &  0.030$\pm$0.027 & 0.74 & 14.5 & $208_{-103}^{+174}$  & - &   \\
(163899) 2003 SD220& 17.63 & 0.791$\pm$0.025 & 0.340$\pm$0.042 & 2.20 & 285.  & $1506_{-467}^{+564}$  & - &    \\
(185851) 2000 DP107& 18.33 & 0.822$\pm$0.184  & 0.137$\pm$0.033  & 0.18  &  2.775 & $445_{-173}^{+369}$  & $54_{-25}^{+35}$ &    \\
(242191) 2003 NZ6 & 19.02 & 0.370$\pm$0.031  & 0.334$\pm$0.070 & 1.50 & 13.531 & $360_{-121}^{+183}$  & - &    \\
(297418) 2000 SP43 & 18.52  & 0.407$\pm$0.019  & 0.388$\pm$0.053 & 1.10 & 6.314 & $186_{-64}^{+122}$  &  - &  \\
(337248) 2000 RH60& 20.11 & 0.846$\pm$0.078 &  0.025$\pm$0.008  & 0.35 & 25.2 & $783_{-288}^{+388}$  & - &   \\
(363027) 1998 ST27 & 19.64  & 0.690$\pm$0.250  & 0.06$\pm$0.07    & 0.10  &  3.0   & $68_{-34}^{+85}$  &  - &   \\
(363505) 2003 UC20& 18.52  & 1.876$\pm$0.037  & 0.029$\pm$0.006 & 0.90  &  29.6 & $489_{-191}^{+296}$  &  - &  \\
(363599) 2004 FG11 & 21.04  & 0.152$\pm$0.003  & 0.306$\pm$0.050  & 0.30  & 7.021  & $191_{-73}^{+137}$  & - &    \\
(385186) 1994 AW1 & 17.65& 0.811$\pm$0.021  & 0.223$\pm$0.045  & 0.20 & 2.519  & $274_{-68}^{+112}$  & $85_{-23}^{+27}$  &    \\
(398188) Agni     & 19.33 & 0.462$\pm$0.006 & 0.137$\pm$0.024 & 1.10 & 21.99 & $321_{-140}^{+188}$  & - &    \\
(410777) 2009 FD  & 22.18  & 0.472$\pm$0.045 & 0.010$\pm$0.003 & 0.46 & 5.8 & $200_{-64}^{+76}$  & -  &  \\
(422686) 2000 AC6 & 21.63 & 0.176$\pm$0.005  & 0.143$\pm$0.021 & 0.30 & 2.444 & $887_{-341}^{+453}$  & $73_{-23}^{+42}$ &   \\
(441987) 2010 NY65 & 21.35  & 0.228$\pm$0.012  & 0.071$\pm$0.014  & 0.30  &  4.971 & $48_{-16}^{+39}$  &  -  &  \\
(443880) 2001 UZ16& 19.38 & 0.25$\pm$0.04 & 0.50$\pm$0.17 & 1.00 & 13.719 & $484_{-140}^{+266}$  & - &   \\
(511684) 2015 BN509& 20.80 & 0.315$\pm$0.065 & 0.093$\pm$0.044 & 1.00 & 5.671 & $190_{-80}^{+128}$  & - &    \\
~~-~~~~2010 VK139  & 23.71 & 0.055$\pm$0.010 & 0.196$\pm$0.114 & 0.50 & 0.030 & $213_{-107}^{+225}$  & - &   \\
\hline
    \end{tabular}
    \label{tab:new_TI}
    
\tablenotetext{~The values of absolute magnitudes $H$ are from the JPL, diameters $D$ and albedos $p_V$ are from \citet{2019PDSS..251.....M}, except for (3200) Phaethon \citep{2019P&SS..167....1T}, (3908) Nyx \citep{2002Icar..158..379B}, (4034) Vishnu and (4660) Nereus \citep{2003Icar..166..116D}, (6489)	Golevka \citep{hudson-etal_2000}, (85990) 1999 JV6 \citep{2020PSJ.....1....9M}, (363027) 1998 ST27, (363505) 2003 UC20, and (443880) 2001 UZ16 \citep{2016AJ....152...63N}, and 2010 VK139 \citep{2014ApJ...784..110M}. The values of maximum light curve amplitudes $\Delta m$ and rotation periods $P$ are taken from LCDB \citep{warner-etal_2009}.  }

\end{table*}

\section{Summary and conclusions}
\label{s:summary}

Here, we presented the new Asteroid Thermal Inertia Analyzer (ASTERIA) model and publicly available corresponding software for determining the surface thermal inertia of asteroids. The model allows thermal inertia estimation based mostly on population-modeled input parameters in its basic variant. However, as in general cases, the model may not work well if all physical parameters are population-based, we identified the set of critical parameters, which includes the diameter, albedo, and rotation periods of an object.

The model and code have been validated using results for Bennu and ten other well-characterized near-Earth asteroids, proving that the ASTERIA is a reliable tool for thermal inertia analysis.
Overall, we find good agreement between our results and the literature, with all terms within $1–\sigma$ of each other. 

It should be noted, however, that our model tends to slightly underestimate the thermal inertia in some cases, especially for low thermal inertia below 200~J m$^{-2}$ K$^{-1}$ s$^{-1/2}$ (see Figure~\ref{fig:TI_comparison}). This potential systematic offset may be worth further investigation once a larger comparison set becomes available. 

A general advantage of the ASTERIA model is that it may be applied to smaller asteroids than the thermophysical modeling, because the Yarkovsky effect is more substantial in smaller objects and, therefore, easier to detect. Additionally, thermophysical modeling requires thermal infrared observations and good shape models, which are currently challenging for asteroids below some 100 meters in size.
In this respect, we note that among all the asteroids' thermal inertia estimates available in the literature, the results for the three smallest ones are obtained with our model (see Fig.~\ref{fig:TI_overview}).

Based on the astrometric measurements and the detection of Yarkovsky-induced acceleration of orbital motion, it is primarily independent of the most widely used approach for the asteroid thermal inertia estimations based on thermophysical modeling. As such, the ASTERIA may also serve as a benchmark test to independently verify the results derived from the thermophysical modeling, one of the longstanding challenges of TPM models \citep{2022PSJ.....3...56H}.

As we found that knowledge of a rotation period is highly relevant for the ASTERIA model to work correctly, we provide a list of high-priority targets (Table~\ref{tab:observable_objects}) and encourage the observers to attempt to obtain light curves of those objects. Determining the rotation periods of those asteroids would allow the thermal inertia estimations by the ASTERIA model.

Finally, we have identified a set of 38 NEAs for which all the input parameters critical for the ASTERIA model to work reliably are available and presented the new thermal inertia for those objects.
Among these 38 NEAs, 29 are classified as potentially hazardous asteroids (PHA, Table~\ref{tab:new_NEAs_orb}). It makes our results highly relevant from the planetary defense point of view. Our sample of new thermal inertia estimates also includes 31 sub-km-sized asteroids,
while there are only 17 other literature values in this size range, highlighting the importance of the ASTERIA model for determining the surface thermal inertia of small asteroids.

Among the limitations, we recall that if the input parameters are not well-defined and have considerable uncertainty, this propagates into significant uncertainties of the thermal inertia estimates by our model. An example of this is the asteroid (1865) Cerberus. Though we obtained results formally in agreement with the reference literature estimate, the situation highlights the model's sensitivity to the Yarkovsky drift determination. Also, for likely elongated objects with large light curve amplitudes, such as Cerberus, our non-sphericity scaling could be rough, contributing to overall uncertainty. Nevertheless, we showed that the obtained results are even in these cases compatible with other estimates. Therefore, results for such objects could be helpful but should be interpreted cautiously. Future better characterization of the input parameters will further improve the precision of our results.

Also, our model could be somewhat affected by spatial thermal inertia variation across the surface of an asteroid. In case of significant spatial variations, an additional error of up to about 5\% could be introduced. For instance, in the case of asteroid Bennu, we found up to a 2\% shift in the result due to spatial variation in thermal inertia from the equator to the polar regions.

Applicability of our model is still limited to the near-Earth asteroids, as the Yarkovsky detections are yet unavailable for main-belt asteroids \citep{2023A&A...674A..12T}.

%%\begin{acknowledgments}
%%
\vskip 7mm

\textbf{The authors appreciate the support from the Planetary Society STEP Grant, made possible by the generosity of The Planetary Society' members. BN also acknowledges support from the project PID2021-126365NB-C21 (MCI/AEI/FEDER, UE) and from the Severo Ochoa grant CEX2021-001131-S funded by MCI/AEI/10.13039/501100011033.}

%%\end{acknowledgments}

\bibliography{asteria.bbl}{}
\bibliographystyle{aasjournal}

%% This command is needed to show the entire author+affiliation list when
%% the collaboration and author truncation commands are used.  It has to
%% go at the end of the manuscript.
%\allauthors

%% Include this line if you are using the \added, \replaced, \deleted
%% commands to see a summary list of all changes at the end of the article.
%\listofchanges

\end{document}